# The IceCube Neutrino Observatory IV: Searches for Dark Matter and Exotic Particles

THE ICECUBE COLLABORATION

# IceCube Collaboration Member List

R. Abbasi[28], Y. Abdou[22], T. Abu-Zayyad[33], M. Ackermann[39], J. Adams[16], J. A. Aguilar[28], M. Ahlers[32], M. M. Allen[36], D. Altmann[1], K. Andeen[28,a], J. Auffenberg[38], X. Bai[31,b], M. Baker[28], S. W. Barwick[24], R. Bay[7], J. L. Bazo Alba[39], K. Beattie[8], J. J. Beatty[18,19], S. Bechet[13], J. K. Becker[10], K.-H. Becker[38], M. L. Benabderrahmane[39], S. BenZvi[28], J. Berdermann[39], P. Berghaus[31], D. Berley[17], E. Bernardini[39], D. Bertrand[13], D. Z. Besson[26], D. Bindig[38], M. Bissok[1], E. Blaufuss[17], J. Blumenthal[1], D. J. Boersma[1], C. Bohm[34], D. Bose[14], S. Böser[11], O. Botner[37], A. M. Brown[16], S. Buitink[14], K. S. Caballero-Mora[36], M. Carson[22], D. Chirkin[28], B. Christy[17], F. Clevermann[20], S. Cohen[25], C. Colnard[23], D. F. Cowen[36,35], A. H. Cruz Silva[39], M. V. D'Agostino[7], M. Danninger[34], J. Daughhetee[5], J. C. Davis[18], C. De Clercq[14], T. Degner[11], L. Demirörs[25], F. Descamps[22], P. Desiati[28], G. de Vries-Uiterweerd[22], T. DeYoung[36], J. C. Díaz-Vélez[28], M. Dierckxsens[13], J. Dreyer[10], J. P. Dumm[28], M. Dunkman[36], J. Eisch[28], R. W. Ellsworth[17], O. Engdegård[37], S. Euler[1], P. A. Evenson[31], O. Fadiran[28], A. R. Fazely[6], A. Fedynitch[10], J. Feintzeig[28], T. Feusels[22], K. Filimonov[7], C. Finley[34], T. Fischer-Wasels[38], B. D. Fox[36], A. Franckowiak[11], R. Franke[39], T. K. Gaisser[31], J. Gallagher[27], L. Gerhardt[8,7], L. Gladstone[28], T. Glüsenkamp[39], A. Goldschmidt[8], J. A. Goodman[17], D. Góra[39], D. Grant[21], T. Griesel[29], A. Gross[16,23], S. Grullon[28], M. Gurtner[38], C. Ha[36], A. Haj Ismail[22], A. Hallgren[37], F. Halzen[28], K. Han[39], K. Hanson[13,28], D. Heinen[1], K. Helbing[38], R. Hellauer[17], S. Hickford[16], G. C. Hill[28], K. D. Hoffman[17], A. Homeier[11], K. Hoshina[28], W. Huelsnitz[17,c], J.-P. Hülss[1], P. O. Hulth[34], K. Hultqvist[34], S. Hussain[31], A. Ishihara[15], E. Jacobi[39], J. Jacobsen[28], G. S. Japaridze[4], H. Johansson[34], K.-H. Kampert[38], A. Kappes[9], T. Karg[38], A. Karle[28], P. Kenny[26], J. Kiryluk[8,7], F. Kislat[39], S. R. Klein[8,7], J.-H. Köhne[20], G. Kohnen[30], H. Kolanoski[9], L. Köpke[29], S. Kopper[38], D. J. Koskinen[36], M. Kowalski[11], T. Kowarik[29], M. Krasberg[28], T. Krings[1], G. Kroll[29], N. Kurahashi[28], T. Kuwabara[31], M. Labare[14], K. Laihem[1], H. Landsman[28], M. J. Larson[36], R. Lauer[39], J. Lünemann[29], J. Madsen[33], A. Marotta[13], R. Maruyama[28], K. Mase[15], H. S. Matis[8], K. Meagher[17], M. Merck[28], P. Mészáros[35,36], T. Meures[13], S. Miarecki[8,7], E. Middell[39], N. Milke[20], J. Miller[37], T. Montaruli[28,d], R. Morse[28], S. M. Movit[35], R. Nahnhauer[39], J. W. Nam[24], U. Naumann[38], D. R. Nygren[8], S. Odrowski[23], A. Olivas[17], M. Olivo[10], A. O'Murchadha[28], S. Panknin[11], L. Paul[1], C. Pérez de los Heros[37], J. Petrovic[13], A. Piegsa[29], D. Pieloth[20], R. Porrata[7], J. Posselt[38], P. B. Price[7], G. T. Przybylski[8], K. Rawlins[3], P. Redl[17], E. Resconi[23,e], W. Rhode[20], M. Ribordy[25], M. Richman[17], J. P. Rodrigues[28], F. Rothmaier[29], C. Rott[18], T. Ruhe[20], D. Rutledge[36], B. Ruzybayev[31], D. Ryckbosch[22], H.-G. Sander[29], M. Santander[28], S. Sarkar[32], K. Schatto[29], T. Schmidt[17], A. Schönwald[39], A. Schukraft[1], A. Schultes[38], O. Schulz[23,f], M. Schunck[1], D. Seckel[31], B. Semburg[38], S. H. Seo[34], Y. Sestayo[23], S. Seunarine[12], A. Silvestri[24], G. M. Spiczak[33], C. Spiering[39], M. Stamatikos[18,g], T. Stanev[31], T. Stezelberger[8], R. G. Stokstad[8], A. Stössl[39], E. A. Strahler[14], R. Ström[37], M. Stüer[11], G. W. Sullivan[17], Q. Swillens[13], H. Taavola[37], I. Taboada[5], A. Tamburro[33], A. Tepe[5], S. Ter-Antonyan[6], S. Tilav[31], P. A. Toale[2], S. Toscano[28], D. Tosi[39], N. van Eijndhoven[14], J. Vandenbroucke[7], A. Van Overloop[22], J. van Santen[28], M. Vehring[1], M. Voge[11], C. Walck[34], T. Waldenmaier[9], M. Wallraff[1], M. Walter[39], Ch. Weaver[28], C. Wendt[28], S. Westerhoff[28], N. Whitehorn[28], K. Wiebe[29], C. H. Wiebusch[1], D. R. Williams[2], R. Wischnewski[39], H. Wissing[17], M. Wolf[23], T. R. Wood[21], K. Woschnagg[7], C. Xu[31], D. L. Xu[2], X. W. Xu[6], J. P. Yanez[39], G. Yodh[24], S. Yoshida[15], P. Zarzhitsky[2], M. Zoll[34]

[1] III. Physikalisches Institut, RWTH Aachen University, D-52056 Aachen, Germany
[2] Dept. of Physics and Astronomy, University of Alabama, Tuscaloosa, AL 35487, USA
[3] Dept. of Physics and Astronomy, University of Alaska Anchorage, 3211 Providence Dr., Anchorage, AK 99508, USA
[4] CTSPS, Clark-Atlanta University, Atlanta, GA 30314, USA
[5] School of Physics and Center for Relativistic Astrophysics, Georgia Institute of Technology, Atlanta, GA 30332, USA
[6] Dept. of Physics, Southern University, Baton Rouge, LA 70813, USA
[7] Dept. of Physics, University of California, Berkeley, CA 94720, USA
[8] Lawrence Berkeley National Laboratory, Berkeley, CA 94720, USA
[9] Institut für Physik, Humboldt-Universität zu Berlin, D-12489 Berlin, Germany
[10] Fakultät für Physik & Astronomie, Ruhr-Universität Bochum, D-44780 Bochum, Germany
[11] Physikalisches Institut, Universität Bonn, Nussallee 12, D-53115 Bonn, Germany
[12] Dept. of Physics, University of the West Indies, Cave Hill Campus, Bridgetown BB11000, Barbados
[13] Université Libre de Bruxelles, Science Faculty CP230, B-1050 Brussels, Belgium
[14] Vrije Universiteit Brussel, Dienst ELEM, B-1050 Brussels, Belgium
[15] Dept. of Physics, Chiba University, Chiba 263-8522, Japan
[16] Dept. of Physics and Astronomy, University of Canterbury, Private Bag 4800, Christchurch, New Zealand
[17] Dept. of Physics, University of Maryland, College Park, MD 20742, USA
[18] Dept. of Physics and Center for Cosmology and Astro-Particle Physics, Ohio State University, Columbus, OH 43210, USA
[19] Dept. of Astronomy, Ohio State University, Columbus, OH 43210, USA
[20] Dept. of Physics, TU Dortmund University, D-44221 Dortmund, Germany
[21] Dept. of Physics, University of Alberta, Edmonton, Alberta, Canada T6G 2G7
[22] Dept. of Physics and Astronomy, University of Gent, B-9000 Gent, Belgium
[23] Max-Planck-Institut für Kernphysik, D-69177 Heidelberg, Germany
[24] Dept. of Physics and Astronomy, University of California, Irvine, CA 92697, USA
[25] Laboratory for High Energy Physics, École Polytechnique Fédérale, CH-1015 Lausanne, Switzerland
[26] Dept. of Physics and Astronomy, University of Kansas, Lawrence, KS 66045, USA
[27] Dept. of Astronomy, University of Wisconsin, Madison, WI 53706, USA
[28] Dept. of Physics, University of Wisconsin, Madison, WI 53706, USA
[29] Institute of Physics, University of Mainz, Staudinger Weg 7, D-55099 Mainz, Germany
[30] Université de Mons, 7000 Mons, Belgium
[31] Bartol Research Institute and Department of Physics and Astronomy, University of Delaware, Newark, DE 19716, USA
[32] Dept. of Physics, University of Oxford, 1 Keble Road, Oxford OX1 3NP, UK
[33] Dept. of Physics, University of Wisconsin, River Falls, WI 54022, USA
[34] Oskar Klein Centre and Dept. of Physics, Stockholm University, SE-10691 Stockholm, Sweden
[35] Dept. of Astronomy and Astrophysics, Pennsylvania State University, University Park, PA 16802, USA
[36] Dept. of Physics, Pennsylvania State University, University Park, PA 16802, USA
[37] Dept. of Physics and Astronomy, Uppsala University, Box 516, S-75120 Uppsala, Sweden
[38] Dept. of Physics, University of Wuppertal, D-42119 Wuppertal, Germany
[39] DESY, D-15735 Zeuthen, Germany
[a] now at Dept. of Physics and Astronomy, Rutgers University, Piscataway, NJ 08854, USA
[b] now at Physics Department, South Dakota School of Mines and Technology, Rapid City, SD 57701, USA
[c] Los Alamos National Laboratory, Los Alamos, NM 87545, USA
[d] also Sezione INFN, Dipartimento di Fisica, I-70126, Bari, Italy
[e] now at T.U. Munich, 85748 Garching & Friedrich-Alexander Universität Erlangen-Nürnberg, 91058 Erlangen, Germany
[f] now at T.U. Munich, 85748 Garching, Germany
[g] NASA Goddard Space Flight Center, Greenbelt, MD 20771, USA

## Acknowledgments

We acknowledge the support from the following agencies: U.S. National Science Foundation-Office of Polar Programs, U.S. National Science Foundation-Physics Division, University of Wisconsin Alumni Research Foundation, the Grid Laboratory Of Wisconsin (GLOW) grid infrastructure at the University of Wisconsin - Madison, the Open Science Grid (OSG) grid infrastructure; U.S. Department of Energy, and National Energy Research Scientific Computing Center, the Louisiana Optical Network Initiative (LONI) grid computing resources; National Science and Engineering Research Council of Canada; Swedish Research Council, Swedish Polar Research Secretariat, Swedish National Infrastructure for Computing (SNIC), and Knut and Alice Wallenberg Foundation, Sweden; German Ministry for Education and Research (BMBF), Deutsche Forschungsgemeinschaft (DFG), Research Department of Plasmas with Complex Interactions (Bochum), Germany; Fund for Scientific Research (FNRS-FWO), FWO Odysseus programme, Flanders Institute to encourage scientific and technological research in industry (IWT), Belgian Federal Science Policy Office (Belspo); University of Oxford, United Kingdom; Marsden Fund, New Zealand; Japan Society for Promotion of Science (JSPS); the Swiss National Science Foundation (SNSF), Switzerland; D. Boersma acknowledges support by the EU Marie Curie IRG Program; A. Groß acknowledges support by the EU Marie Curie OIF Program; J. P. Rodrigues acknowledges support by the Capes Foundation, Ministry of Education of Brazil; A. Schukraft acknowledges the support by the German Telekom Foundation; N. Whitehorn acknowledges support by the NSF Graduate Research Fellowships Program.

# Indirect search for Solar dark matter with AMANDA and IceCube

THE ICECUBE COLLABORATION[1]
[1] *See special section in these proceedings*

**Abstract:** It has been suggested that the Milky Way dark matter halo consists of Weakly Interactive Massive Particles (WIMPs) that scatter and accumulate in the center of gravitational wells like the Sun. In certain models this provides a source of high energy neutrinos originating in the centre of the Sun that may be detected by a neutrino detector on the Earth. We describe a search for Solar WIMP dark matter using AMANDA/IceCube data from 2008 with the 40-string IceCube detector configuration. We present combined limits using data acquired from 2001-2008 on the WIMP induced muon flux from the Sun for dark matter particle masses between 50 and 5000 GeV. We test WIMP model hypotheses of the lightest MSSM neutralino and the lightest Kaluza-Klein particle.

**Corresponding author:** Olle Engdegård (*olle@fysast.uu.se*)
*Department of physics and astronomy, Uppsala University, S-75120 Uppsala, Sweden*

## 1 Introduction

The intriguing question of why less than one fifth of the matter in our universe seems to be of the ordinary sort that emits electromagnetic radiation and the rest being invisible by all other means than gravitational inference, has been an outstanding problem in physics for almost 80 years. The most popular current hypotheses involve Weakly Interacting Massive Particles (WIMPs) that are distributed in and around galaxies as a halo. In several theories these particles will accumulate in massive objects, like the Sun, by scattering (weakly) multiple times. With each scatter momentum is lost and the particle becomes gravitationally trapped in the centre regions of the object. Should the WIMPs then be Majorana particles they will pair-wise annihilate and, in some channels, give rise to a neutrino signal that is detectable on Earth.

In this paper we present an explicit search for Solar WIMPs described by the Minimal Supersymmetric Standard Model (MSSM) [1], where the lightest supersymmetric particle $\tilde{\chi}_1^0$, henceforth denoted $\chi$, is a promising dark matter candidate. The two most extreme annihilation channels are studied; $\chi \rightarrow b\bar{b}$ producing the softest possible neutrino spectrum, and $\chi \rightarrow W^+W^-$ producing the hardest possible spectrum. Note that for $m_\chi < m_W$ we consider instead the channel $\chi \rightarrow \tau^+\tau^-$.

The same event selection is also applied to search for another potential dark matter candidate, the Lightest Kaluza-Klein Particle (LKP), arising from theories of Universal Extra Dimensions (UED) [2]. The 5-dimensional model we considered here is described by the mass $m_\gamma^{(1)}$ of the LKP, the first photon excitation, and the mass splitting $\Delta_{q^{(1)}} = (m_{q^{(1)}} - m_{\gamma^{(1)}})/m_{\gamma^{(1)}}$, where $m_{q^{(1)}}$ is the mass of the first quark excitation. The Monte Carlo defining the signal here, with $\Delta_{q^{(1)}} = 0$, was taken from [3].

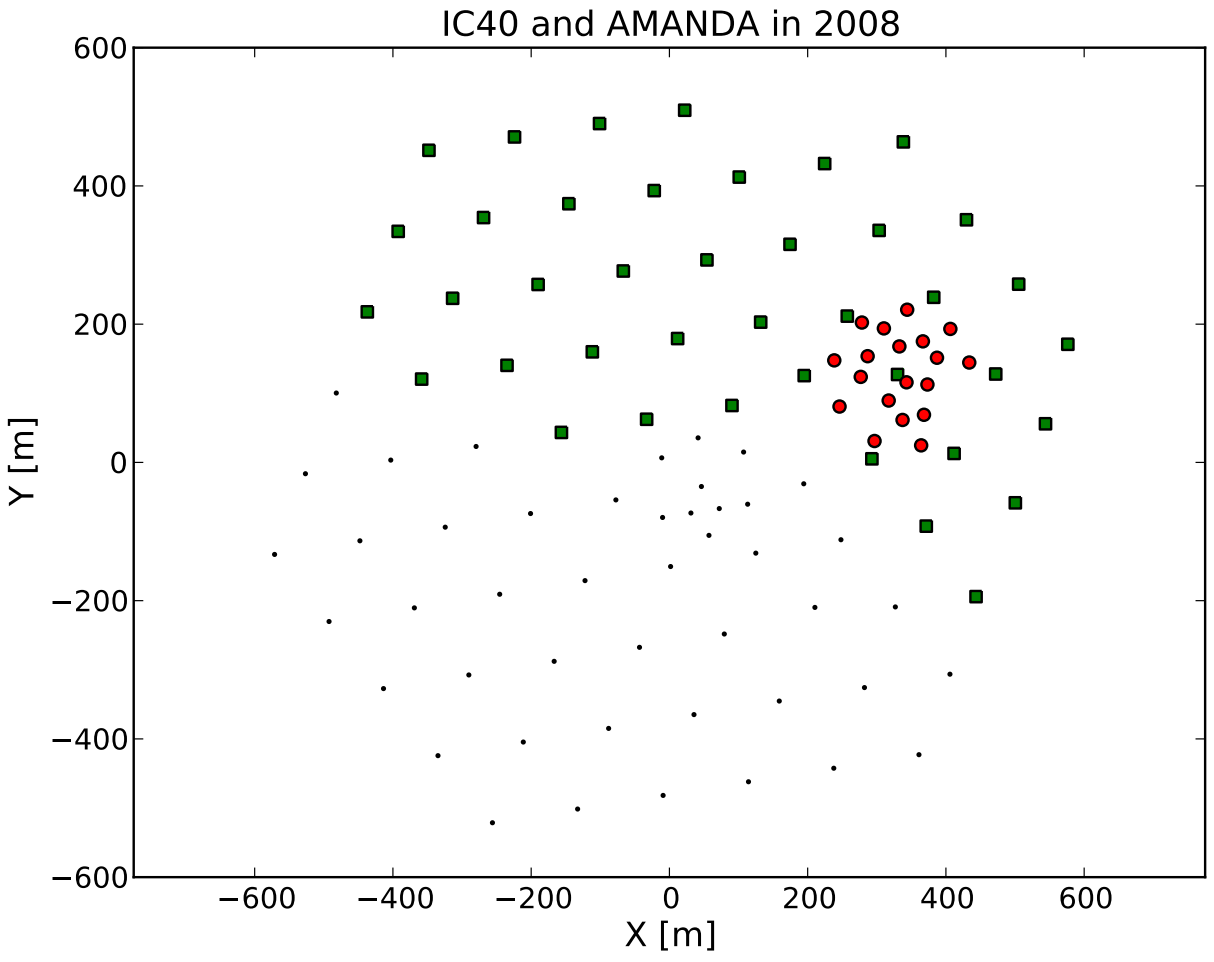

Figure 1: IceCube, depicted from above, in its 40-string configuration (squares) together with AMANDA (circles). The dots show the full 86-string IceCube detector in 2010.

The world's largest neutrino observatory, IceCube [4], was halfway to completion in 2008 with 40 strings, each containing 60 digital optical modules, deployed in the lay-out shown in figure 1. AMANDA, the predecessor detector to IceCube, underwent a low-energy trigger upgrade and integration of its Data Acquisition System (DAQ) with that of IceCube. As seen in Figure 1, AMANDA, with a markedly lower energy threshold than IceCube, was horizontally enclosed by the 2008 IceCube array. Data from this year thus provided a unique opportunity to use the two detectors as one. In particular, IceCube could be used as an efficient veto for horizontal atmospheric muons entering AMANDA. This opportunity was not missed, and we present here the result of a search for Solar dark matter using this data, and follow up by combining the result with previous AMANDA and IceCube limits.

## 2 Data and simulation

### Data

An analysis was undertaken on the 2008 IceCube-AMANDA dataset amounting to a total livetime of 149 days. The livetime considers the period when the sun was below the horizon (March to September). A triggered sample of $1.7\times10^{10}$ events exists in this dataset, the vast majority of which are atmospheric muons.

The chosen event-sample was to be reduced by a factor of $\sim10^{-7}$ using variable cuts and a machine learning algorithm in an effort to maximise the separation of signal and background. When optimising the event selection, described further in section 3, data from November 2008 was used and then discarded.

### Simulation

The signal Monte Carlo was generated with DarkSUSY [5], simulating WIMPs and LKPs for a range of different particle masses. As mentioned in section 1, we generated two MC samples for each MSSM mass corresponding to soft and hard neutrino spectra.

The atmospheric muon background was simulated with CORSIKA [6] using the Hörandel [7] model for the cosmic ray spectrum and composition. The background of atmospheric $\nu_\mu$ was generated with the Honda et al.[8] conventional flux model and the Enberg et al.[9] prompt flux model. After simulating charged particle propagation with MMC [10] and light propagation in the ice with PHOTONICS [11], the detector response was finally simulated using the IceCube simulation software IceSim.

## 3 Event selection

This analysis divided the data into two data streams: One stream used IceCube as the main detector searching for high WIMP masses, using as a signal template the 1000 GeV WIMP model with the hard spectrum. The second stream used AMANDA as the main fiducial volume and IceCube as a sophisticated veto, with the 100 GeV soft-spectrum WIMP as a signal template.

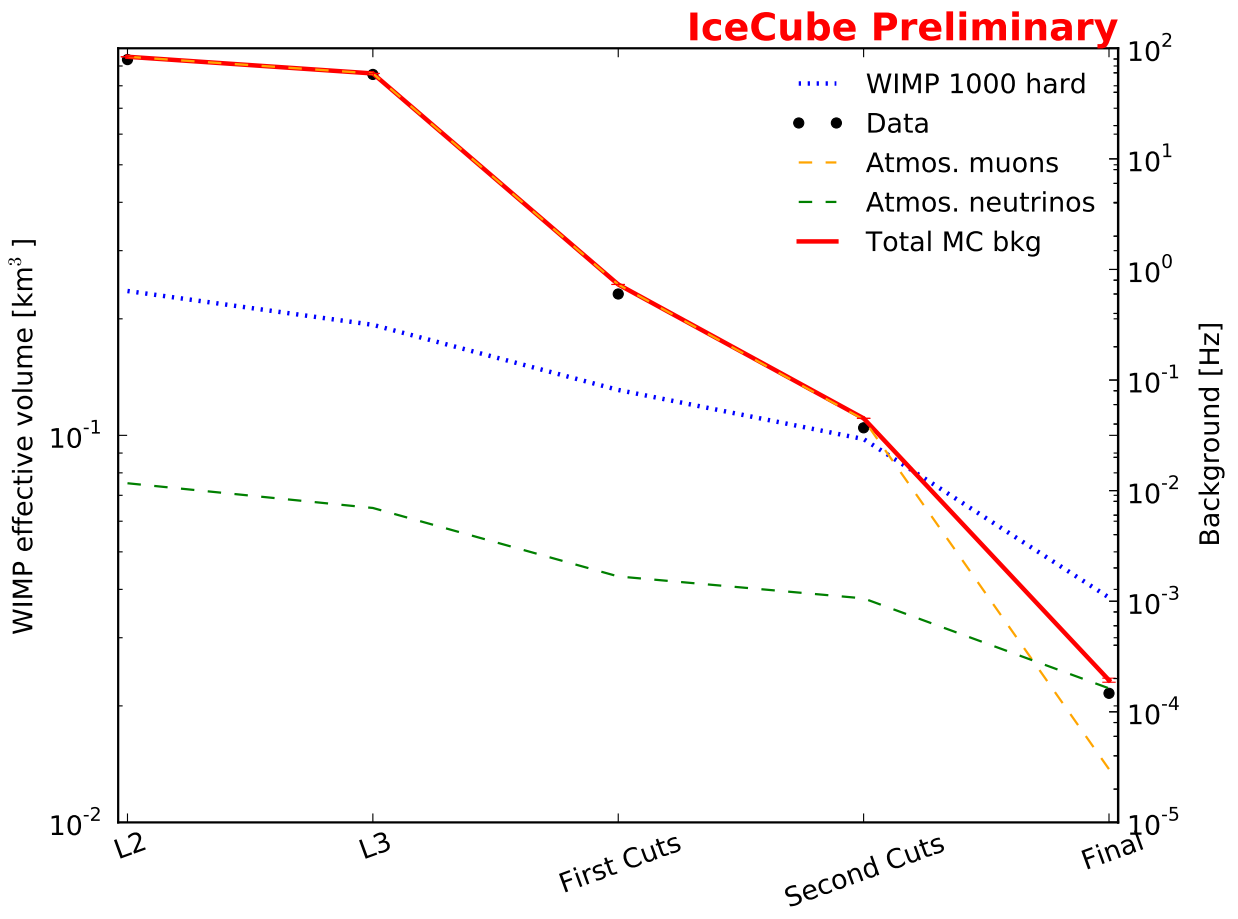

Figure 2: Effective volume for a 1000 GeV WIMP signal on the left y-axis and background rates on the right y-axis, as functions of the accumulated data reduction cuts in the IceCube-stream. In the steep step to the final level a cut was applied on SVM output values.

In each stream a series of linear cuts were applied to data and MC, optimised individually using the signal templates and data from November 2008 as the estimate for the background. In the AMANDA-stream no more than 4 hits were allowed in IceCube, effectively favouring low-energy events. In both streams we made log-likelihood reconstructions of the particle tracks, requiring them to be upward-going and good fits to the hit information. The progression of data reduction is visualised in figure 2.

After applying the linear cuts a Support Vector Machine (SVM) [12] was trained to separate signal from background. A separate signal MC sample and November data as background were again used, the both of which were discarded after the training. When applying a trained SVM on a set of events, a single number $Q \in [0, 1]$ is obtained for each event that characterises it as signal-like or data-like, which can be used as a powerful multivariate cut parameter.

Starting with a set of about 20 variables, an iterative method was used to eliminate variables that did not contribute significantly to the overall separation. In this way, 12 and 10 variables were chosen for the SVMs in the AMANDA- and IceCube-stream respectively, covering a large part of the available phase-space and showing modest internal correlations.

The final cut on $Q$ was set at 0.1, which optimised the *sensitivity* of the dark matter search, i.e., the median 90% upper limit in a large set of random samples drawn from a background distribution without any signal content. It is noted that the neutrino spectra from the Kaluza-Klein signal are very similar to the hard MSSM spectra once convoluted in the propagation through the Sun and processed as IceCube triggers. For this reason, no separate optimisation was necessary to study this potential signal.

### Uncertainties

A number of systematic effects were studied which impact the effective volume of the signal. Uncertainties in the neutrino oscillation parameters, neutrino-nucleon cross-section, muon propagation through ice and calibration of detector time and geometry, all change $V_{\text{eff}}$ less than 5% when individually shifted $1\sigma$ from their nominal values. The uncertainty in light propagation through the ice and the absolute sensitivity of the PMTs remain the dominant systematic effect. This was estimated from dedicated Monte Carlo studies. The total systematic uncertainty, the squared sum of all effects, was calculated for 3 template energy regions: The AMANDA-region, the combined region and the IceCube-region were given the uncertainties 26%, 25% and 15% respectively.

## 4 Search for a Solar excess

At the final cut level, the 2008 data sample contains a total of 3012 events. We examine the properties of the space angles ($\psi_i$) between the reconstructed track and the Sun. The angular resolution, measured as the median space angle for neutrinos in the signal MC sample, is 2.2° for the 1000 GeV hard-spectrum signal and 9.0° for the 100 GeV soft-spectrum signal. The distribution of IceCube data in the direction of the Sun is shown in figure 3 together with the expected background. This background probability density function (PDF) is constructed from actual data using the reconstructed zenith-angles and random, oversampled, azimuth angles.

Since the AMANDA- and IceCube-streams have completely non-overlapping event samples, we will treat them as separate sets of observations, denoted A08 and IC08, that can be combined using techniques described below. In addition to these samples, we also include a sample from an earlier analysis of AMANDA data from 2001-2006 [13] (A01-06, around 2000 events) and the sample from the 22-string IceCube data of 2007 [14] (IC07, 6946 events) to prepare a final analysis that incorporates up to four independent event samples. For all WIMP channels we compare the individual sensitivities and exclude, for computational efficiency, samples yielding a sensitivity 10 times weaker than the strongest. In the results for the Kaluza-Klein models, e.g., A08 did not contribute significantly and was dropped in favour of IC07 and IC08.

The two older samples were both obtained in a similar manner to what is described in this paper. In A01-06 a Boosted Decision Tree (BDT) was trained and optimised for each WIMP channel using 21 variables, in this way performing several parallel searches in a detector livetime of 812 days. The optimal BDT cut was chosen to minimise the number of events needed to have 90% probability of a $5\sigma$ discovery.

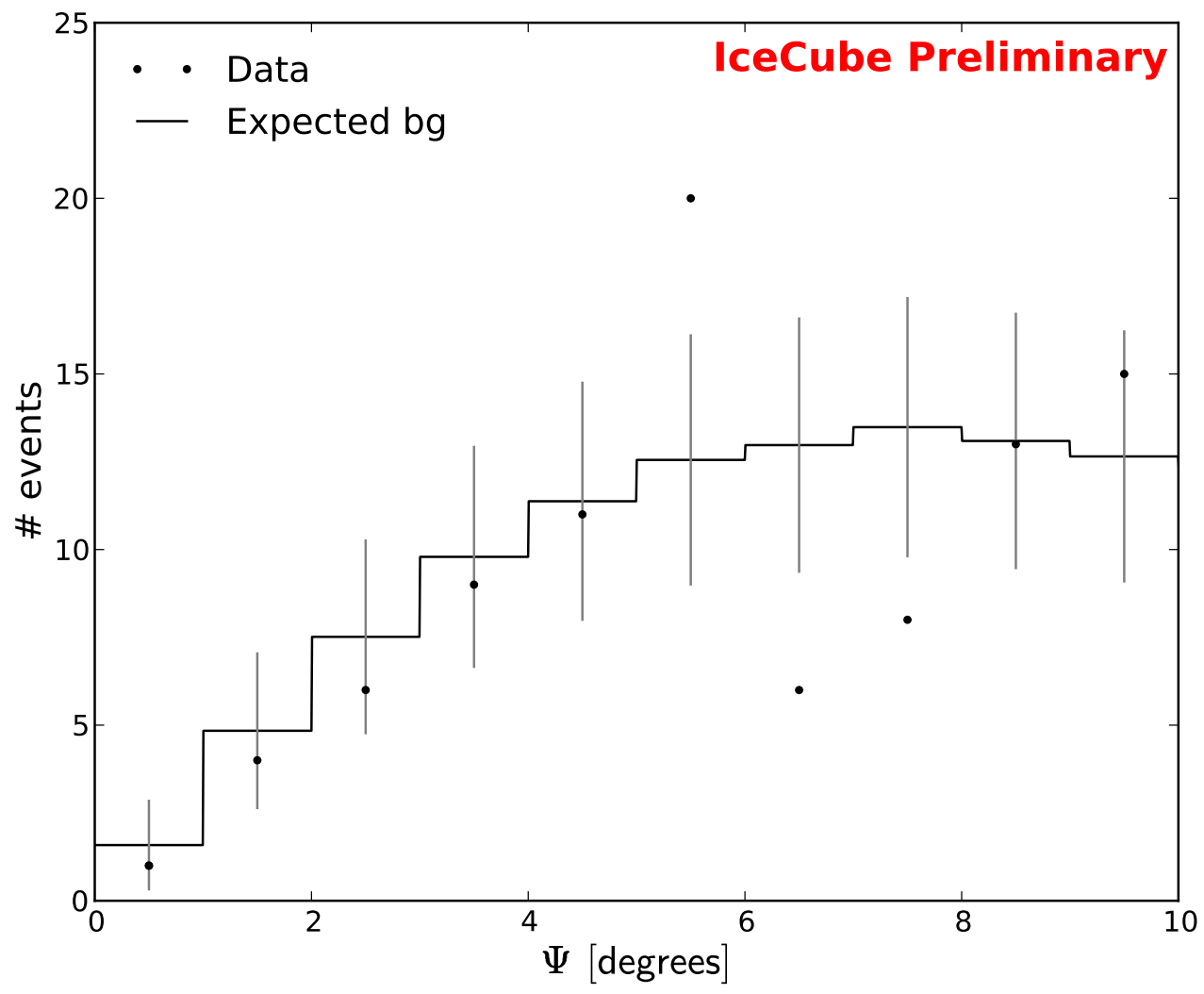

Figure 3: The distribution of space angles between the reconstructed event track and the Sun, for IceCube data at the final cut level and the expected background, in the region near the Sun. The error bars on the background are $1\sigma$ Poisson standard deviations.

The IC07 sample resulted from a combination of straight cuts and two SVMs. Each SVM was trained with 6 different uncorrelated variables, and the optimised cut was made on the product of the two output values.

The goal here is to obtain a confidence interval on the number of signal events, $\mu$, in the total data set. We use the same procedure and profile-likelihood method as outlined in [15], but adapt it for the more general case of $N$ independent experiments. The total likelihood of having $\mu$ signal events in the data set becomes

$$\mathcal{L}(\mu) = \sum_j^N \mathcal{L}_j(\mu_j) = \sum_j^N \prod_i^{n_j^{\text{obs}}} f_j(\psi_i^j|\mu_j), \qquad (1)$$

where each $\mu_j$ is weighted with the livetime and effective volume of the individual experiment using

$$\mu_j = \mu \frac{t_j^{\text{live}} V_j^{\text{eff}}}{\sum_k^N t_k^{\text{live}} V_k^{\text{eff}}}, \qquad (2)$$

so that $\sum_j \mu_j = \mu$. In equation 1, $f(\psi|\mu)$ is the total PDF of expected background plus a signal content of size $\mu$.

Table 1 shows the upper limit, at 90% confidence level, for $\mu$ (or the entire confidence interval if the lower bound is not zero) together with the sensitivity as defined in section 3, the best fit of $\mu$ to data and the probability of observing at least the current deviation from the sensitivity. This was consistent with $\mu = 0$, and the upper edge of the interval was converted to an upper limit, following [15], on the muon flux from the Sun.

| Mass (GeV) | $\mu_{90}$ 90% CI | $\overline{\mu}_{90}$ Sens. | $\mu$ Best fit | P(obs) |
|---|---|---|---|---|
| **Soft** | | | | |
| 50 | 26.0 | 29.2 | 0.0 | 0.43 |
| 100 | 23.7 | 20.3 | 2.8 | 0.37 |
| 250 | [0.6, 27.1] | 13.7 | 10.8 | 0.08 |
| 500 | 12.7 | 17.3 | 0.0 | 0.34 |
| 1000 | 14.0 | 17.1 | 0.0 | 0.38 |
| 3000 | 9.9 | 17.1 | 0.0 | 0.21 |
| 5000 | 9.4 | 17.1 | 0.0 | 0.20 |
| **Hard** | | | | |
| 50 | 20.6 | 16.2 | 3.0 | 0.33 |
| 100 | 19.4 | 12.2 | 4.7 | 0.20 |
| 250 | 15.3 | 16.8 | 0.0 | 0.43 |
| 500 | 10.9 | 16.8 | 0.0 | 0.26 |
| 1000 | 6.8 | 15.9 | 0.0 | 0.13 |
| 3000 | 9.3 | 16.4 | 0.0 | 0.22 |
| 5000 | 8.7 | 15.8 | 0.0 | 0.22 |

Table 1: With $\mu$ as the number of signal events in the final combined sample, the columns show the 90% CI or upper limit, the sensitivity, the best fit to data and the probability to observe at least this deviation from the sensitivity

## 5 Results

Taking into consideration the uncertainties discussed in section , figure 4 shows the 90% CL upper limits on the muon flux from the Sun by self-annihilating dark matter modelled by MSSM and Kaluza-Klein. In the MSSM plot we outline the MSSM model space not yet excluded by the direct detection experiments CDMS [16] and XENON [17], generated by an extensive parameter scan. The lower figure indicates the available Kaluza-Klein model space for considered values of $\Delta_{q^{(1)}}$. We refer the reader to [3] for a more detailed account of probing the Kaluza-Klein space with IceCube.

## 6 Discussion

We have presented an indirect search for Solar WIMP dark matter using data taken with the integrated AMANDA/IceCube detector in 2008. Combining this data with two other event samples from AMANDA and IceCube from 2001-2007 we establish the strongest current upper limit on the muon flux from WIMP dark matter annihilation in the Sun. The completed IceCube detector has now started to take data and, together with the DeepCore low-energy extension, the expected sensitivity will probe substantial regions of the previously untested WIMP parameter space.

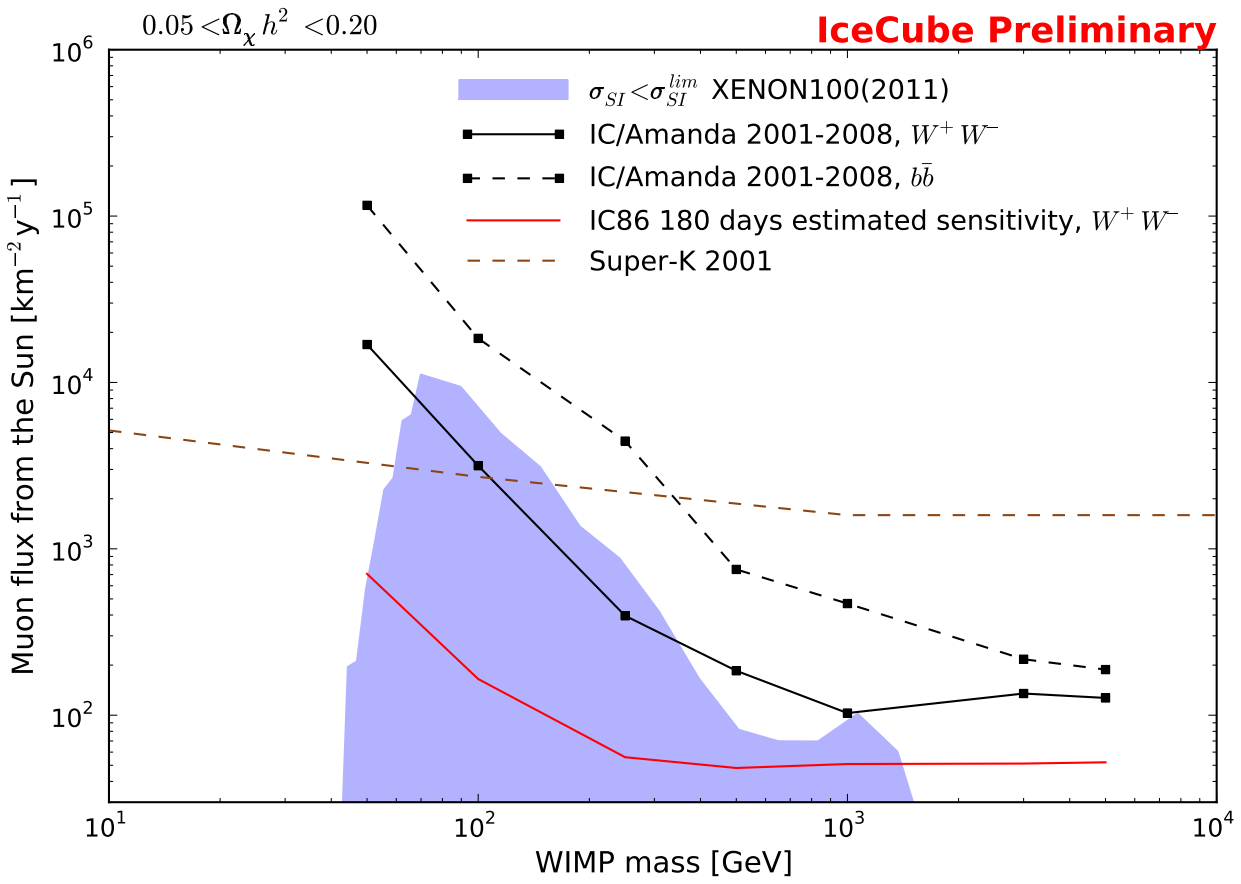

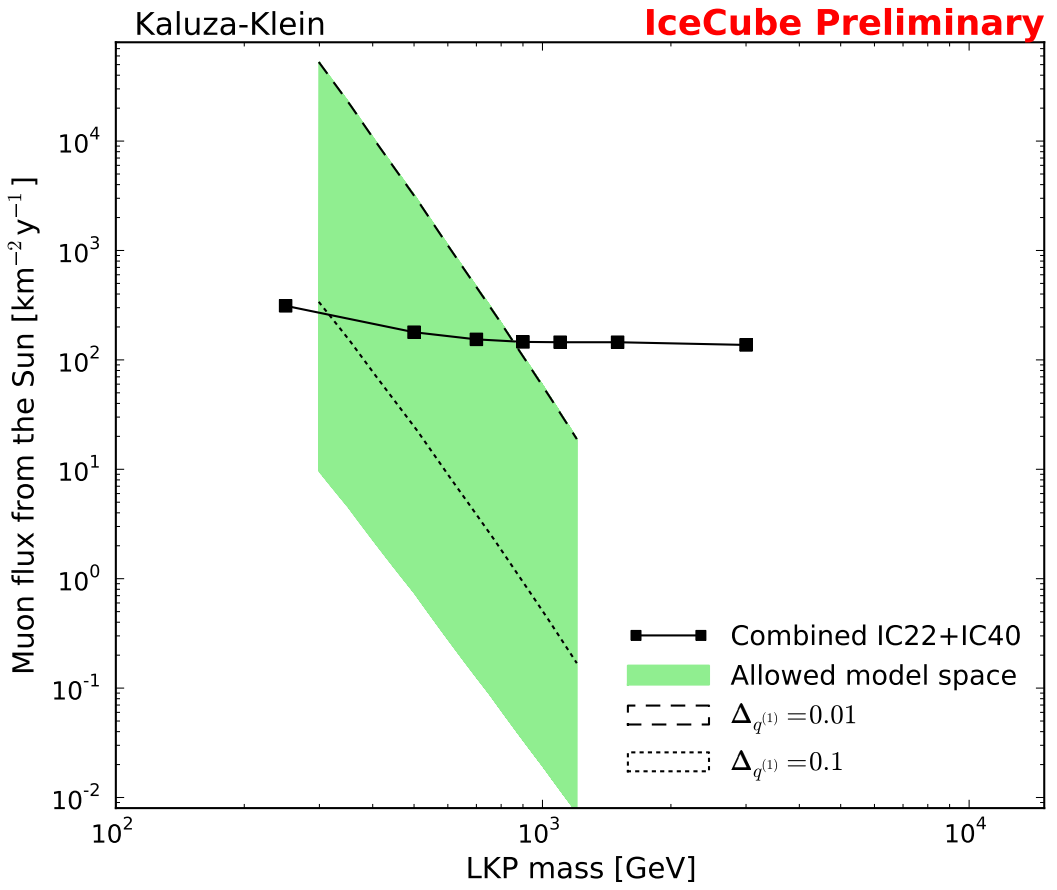

Figure 4: Limits on the WIMP induced muon flux from the Sun modelled by MSSM (above) and Kaluza-Klein (below). In the MSSM limit labeled $W^+W^-$, the channel $\tau^+\tau^-$ was used for $m_\chi < m_W$.

# Searches for Dark Matter Annihilations in the Sun with IceCube and DeepCore in the 79-string configuration

THE ICECUBE COLLABORATION[1]
[1] *See special section in these proceedings*

**Abstract:** Dark matter could be indirectly detected through the observation of neutrinos produced as part of its self-annihilation process. Possible signatures are an excess neutrino flux from the Sun, where dark matter could be gravitationally trapped. The recent commissioning of the full DeepCore sub-array, a low-energy extension of the IceCube neutrino observatory, offers exciting opportunities for neutrino physics in the energy region of 10 GeV to 1 TeV. DeepCore's improved energy reach will, in particular, provide sensitivity to neutrinos from attractive WIMP candidates, like the neutralino or the lightest Kaluza-Klein particle (LKP), down to WIMP masses in the region of about 50 GeV. This will lead to stringent upper limits on the flux of muons from dark matter annihilations in the Sun and constraints on WIMP-proton cross-sections. We report on the search for dark matter annihilations in the Sun with the IceCube neutrino detector in the 79-string configuration, which includes the full DeepCore sub-array. Furthermore, we review the impacts of IceCube observations of the Sun upon supersymmetric (SUSY) dark matter models.

**Corresponding authors:** M. Danninger[2] (*danning@fysik.su.se*) and E. Strahler[3] (*erik.strahler@vub.ac.be*)
[2] *Department of Physics, Stockholm University, AlbaNova, S-10691 Stockholm, Sweden*
[3] *Vrije Universiteit Brussel, Dienst ELEM, B-1050 Brussels, Belgium*

## 1 Introduction

While the presence of dark matter in the universe has been inferred through its gravitational interactions, it has yet to be directly or indirectly observed. One of the most promising and experimentally accessible candidates for dark matter are so-called Weakly Interacting Massive Particles (WIMPs). In theories of the Minimal Supersymmetric Standard Model (MSSM), the WIMP can take the form of the lightest neutralino, $\chi$ [1], while in the framework of universal extra dimensions (UED) [2] it would be the lightest Kaluza-Klein (KK) particle (LKP). Here, we consider the first excitation of the KK photon, $\gamma^{(1)}$, to be the LKP. In current models, WIMPs are predicted to have a mass in the range of a few 10's of GeV to a few TeV. Whatever their underlying physics, these WIMPs may be swept up by the Sun on its transit through the galactic halo and become gravitationally bound by scattering weakly on solar nucleons. Over time, this leads to an accumulation of dark matter in the center of the Sun, exceeding the mean galactic density. Self-annihilation to standard model particles may result in a flux of high energy neutrinos that will be spectrally dependent on the annihilation channel and WIMP mass, and which can be searched for as a point-like source with neutrino telescopes such as IceCube. Several analyses [3, 4, 5] performed on the data from the IceCube detectors have already set stringent limits on the WIMP induced muon flux from the Sun and the spin-dependent WIMP-proton scattering cross section comparable to or better than those of direct detection experiments.

### 1.1 The IceCube Telescope

The IceCube Neutrino Telescope [6] records Cherenkov light in the ice from relativistic charged particles created in neutrino interactions in the vicinity of the detector. By recording the arrival times and intensities of these photons with optical sensors, the direction and energy of the muon and parent neutrino may be reconstructed. IceCube instruments 1 $km^3$ of glacial ice at the South Pole with 5160 Digital Optical Modules (DOMs) on 86 strings deployed between depths of 1450m-2450m. Eight more densely instrumented strings optimized for low energies plus the 12 adjacent standard strings at the center of the detector geometry make up the DeepCore subarray, which increases the sensitivity at low energies and substantially lowers the energy threshold. In addition, by using the surrounding IceCube strings as a veto, DeepCore will enable searches at low energies in the southern hemisphere, transforming IceCube into a full sky observatory. The current analysis of data uses the 79 string configuration of the detector, including 6 DeepCore strings, marked with squares in Figure 1.

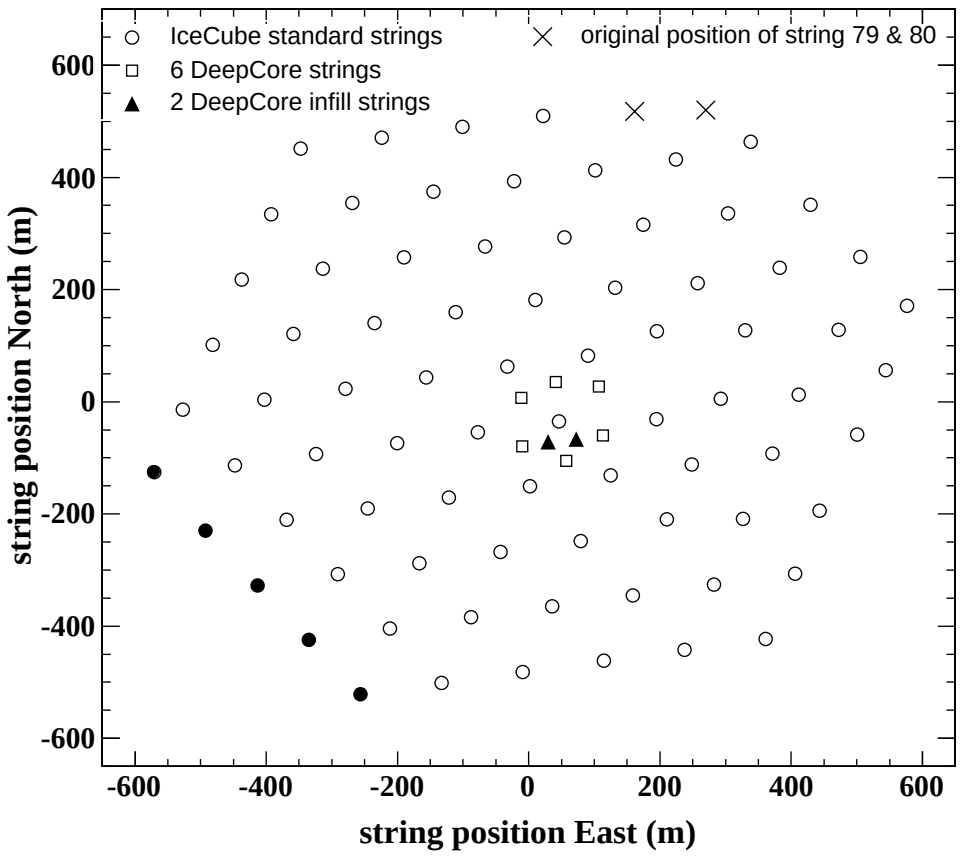

Figure 1: Top view of the final deployed IceCube-86 detector configuration. Standard IceCube strings are shown in circles, initial DeepCore strings in squares, 'in-fill'-strings in DeepCore in triangles and their original design position with crosses. All strings without solid markers make up the 79 string configuration.

The sensitivity study we present for the full 86 string detector uses the initially proposed DeepCore geometry consisting of the same 6 additional strings, as present in the 79 string detector.

## 2 86 String Sensitivity Study

In order to gauge the dark matter physics potential of IceCube as well as the impact of the DeepCore subarray, we have performed a detailed study to determine the sensitivity of the 86 string detector to signals originating from dark matter annihilations in the center of the Sun. (note, DeepCore infill strings, as indicated in Fig. 1, are assumed at their original design position within this study.) Contrary to previous estimates, this study was performed as a full analysis in all details, including detailed data processing and event selection, while making conservative choices where possible. This gives us a realistic expectation of the capabilities of IceCube to observe dark matter induced signals, given the state of data extraction, reconstruction, and signal discrimination techniques available at the time of the study, conducted prior to the start of 79 string data taking.

### 2.1 Data Sample

High statistics atmospheric muon and neutrino background samples are simulated including both single events as well as multiple muon coincidences within the detector. The propagation of muons and photons in the ice is simulated [7, 8] taking measured ice properties into account [9]. All signal simulations use `DarkSUSY` [10] and `WimpSim` [11].

The primary WIMP annihilation spectrum is extremely different between $\chi$ and $\gamma^{(1)}$. For the LKP, branching ratios are fixed from theory, and therefore a 'true' spectrum can be simulated, whereas every MSSM model will result in a different set of branching ratios. To cover the full range, we assume $100\%$ branching into the two end points of the spectrum, i.e the 'soft' $b\bar{b}$ and 'hard' $W^+W^-$ ($\tau^+\tau^-$ below $80.2$ GeV) channels. WIMP signals are generated for 7 different WIMP masses with the 'hard', 'soft' and 'LKP' spectra. The selected masses for the Neutralino models range from $50$ GeV to $5000$ GeV and for the LKP models between $300$ GeV to $1500$ GeV. Three different algorithms are used to trigger the detector; the standard IceCube simple majority trigger 'SMT' with multiplicity threshold $8$, a string trigger ($5$ hit DOMs within $7$ adjacent DOMs on the same string within $1000$ ns) and a lower threshold SMT trigger for the fiducial DeepCore volume optimized for low energies.

### 2.2 Event Selection

After performing likelihood-based reconstructions to achieve the best possible estimate of the muon direction, a series of increasingly stringent event selections are applied to remove the cosmic-ray induced backgrounds. Events are selected which appear to originate from the direction of the Sun and which exhibit low energy signatures. These cuts are not optimized for each model individually, but rather are chosen to provide good general signal retention, ranging from $25\%$ to $70\%$, while rejecting as much background as possible ($\sim 4 \times 10^{-3}$). The remaining signal and background is separated through the use of multivariate training algorithms. In order to determine the best possible approach, multiple algorithms were tested, including support vector machines (SVM), neural networks (NN), and boosted decision trees (BDT). It was found that the best sensitivity could be achieved through the combined use of a BDT and a NN applied on independent sets of observables with low correlation and high discrimination power. In order to further maximize the sensitivity across all models, and taking into account the fact that observables are often strikingly different in different energy regimes, these machines were trained for 2 classes of models, 'high energy' and 'low energy', depending on the resulting muon spectrum after dark matter annihilation and neutrino interaction in the ice. A selection was made on the product of the outputs of the machines to maximize the Model Rejection Factor (MRF) [12] of the analysis. This selection was then tightened in order to achieve a conservative result in light of the fact that the study is limited by statistics on the background simulation. With respect to the straight cuts level, the background reduction is $1 \times 10^{-5}$ with signal efficiencies ranging within $20\%$ to $30\%$.

From the two final 'low energy' and 'high energy' event selections of the signal simulation, we derive the effective area for muon neutrinos from the direction of the Sun as a function of neutrino energy, see Fig. 2. Also shown in the figure is the median angular error between the reconstructed muon and the neutrino directions, Θ. The result is an average over the austral winter, during which the Sun is below the horizon.

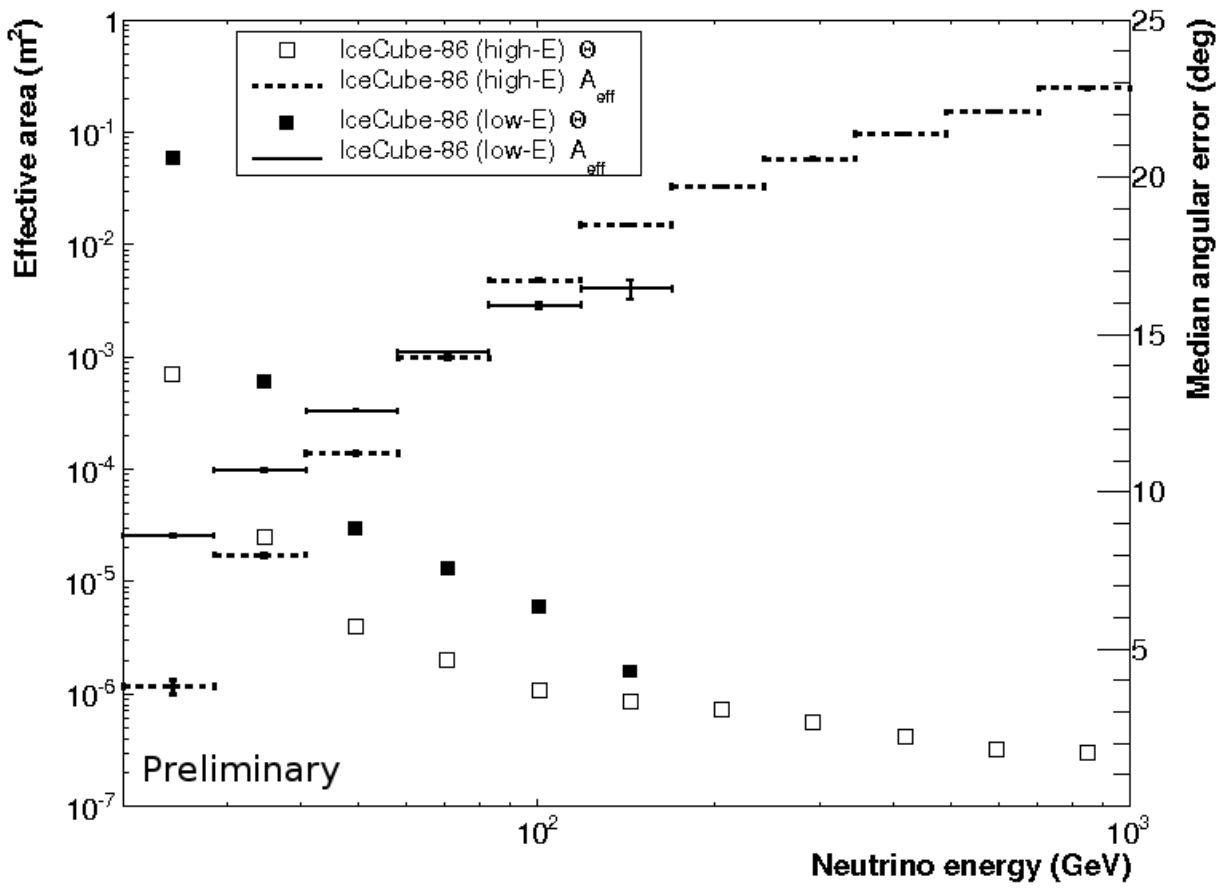

Figure 2: Effective area and median angular error for the final event selection as a function of neutrino energy in the range 20−1000 GeV, for muon neutrinos from the direction of the Sun for the low and high energy cut selection. The result is an average over the austral winter.

### 2.3 Sensitivity

After final event selection, the atmospheric muon background reduction is $< 5.2 \cdot 10^{-8}$, which implies that surviving events in the background sample are dominated by atmospheric neutrinos. We search for an excess of neutrino events over this expected background from the direction of the Sun in a specifically determined search cone, the angular size of which is determined by maximizing the MRF= $\bar{\mu}_s^{90\%}/n_s$, where $n_s$ is the number of simulated signal events remaining in the search cone, and $\bar{\mu}_s^{90\%}$ is the average Feldman-Cousins $90\%$ confidence level upper limit on the signal given an expected background $n_{\rm bg}$ and no true signal content. [13] We derive this $90\%$ C.L. average upper limit, or sensitivity, for the 86 string detector configuration assuming a livetime of 180 days, and from this calculate the limit on the neutrino to muon conversion rate $\Gamma_{\nu\rightarrow\mu}$. Using the signal simulation, we can convert this rate to a limit on the WIMP annihilation rate in the Sun, $\Gamma_{\rm A}$.

Assuming equilibrium between WIMP capture and annihilation in the Sun, the capture rate is then directly proportional to the WIMP-proton scattering cross sections $\sigma_{SI}$ and $\sigma_{SD}$. By assuming that the capture is dominated either by spin-independent or spin-dependent scattering, it is possible to derive sensitivities for $\sigma_{SI}$ and $\sigma_{SD}$ [14]. Results on $\sigma_{SD}$ are shown in Figure 3 for the case of the LKP and the neutralino, and compared to other experimental limits.

## 3 79 String Data Analysis

Data taking with the 79 string configuration of IceCube (including 6 DeepCore strings) took place between May of 2010 and May of 2011, and analysis of this dataset is underway. Several improvements have been realized since the completion of the study detailed above and we describe their potential impact on the sensitivity of the analysis here. While this work is ongoing, it is anticipated that results will be seen in the coming months.

• The low energy trigger in actual data taking was reduced from the planned 4 required hit DOMs within the DeepCore fiducial volume to a less stringent 3 hit DOMs. This further extends the acceptance of the lowest energy events, and boosts the sensitivity for low mass WIMPs.

• In previous datasets, hit information was recorded only for those DOMs which had a timing coincidence with their vertical neighbors. This virtually eliminates random noise hits, but at the same time removes a not insubstantial fraction of real physics hits. Beginning in 2009, this condition is relaxed, and limited hit information is read out for all DOMs which see light in a triggered event. Noise cleaning algorithms have been developed to clean this expanded dataset of unwanted noise while retaining physics information vital for low energy analysis.

• Rather than considering only those events in a search cone around the angular position of the Sun, it is possible to conduct the search utilizing a more sophisticated likelihood technique. Probability distribution functions (PDFs) for WIMP signal and $\nu_{atm}$ background can be constructed relative to the solar position. The signal PDF, $f_s(\Psi)$, is well known for each model from simulation, while the background PDF, $f_{bg}(\Psi)$, can be found by randomizing the azimuth angle in the final event sample. The likelihood that an event belongs to either of the PDFs can be evaluated, and the most likely signal contribution in the final event sample determined. This method is superior to a search cone in that all events are considered, while still penalizing those that lie further away from the Sun's location.

• Extending searches to the southern hemisphere opens up the entire period when the Sun would otherwise be above the horizon, or the austral summer. This doubles the potential livetime of the experiment, though the challenges in reducing the downgoing atmopsheric muon backgrounds are immense. Much of this background can be removed by considering DeepCore and its surrounding strings as a fiducial volume and using the surrounding IceCube strings as an active veto. Furthermore, techniques have been developed to determine the likelihood that a reconstructed muon originates within the detector volume, a sign that it was the result of a neutrino interacting in the ice.

Taking these and other improvements into account, we expect to be able to substantially improve the sensitivity of the analysis over the baseline described in the study above, most notably at the lowest energies, where background dis-crimination is difficult and well reconstructed signal event rates are low.

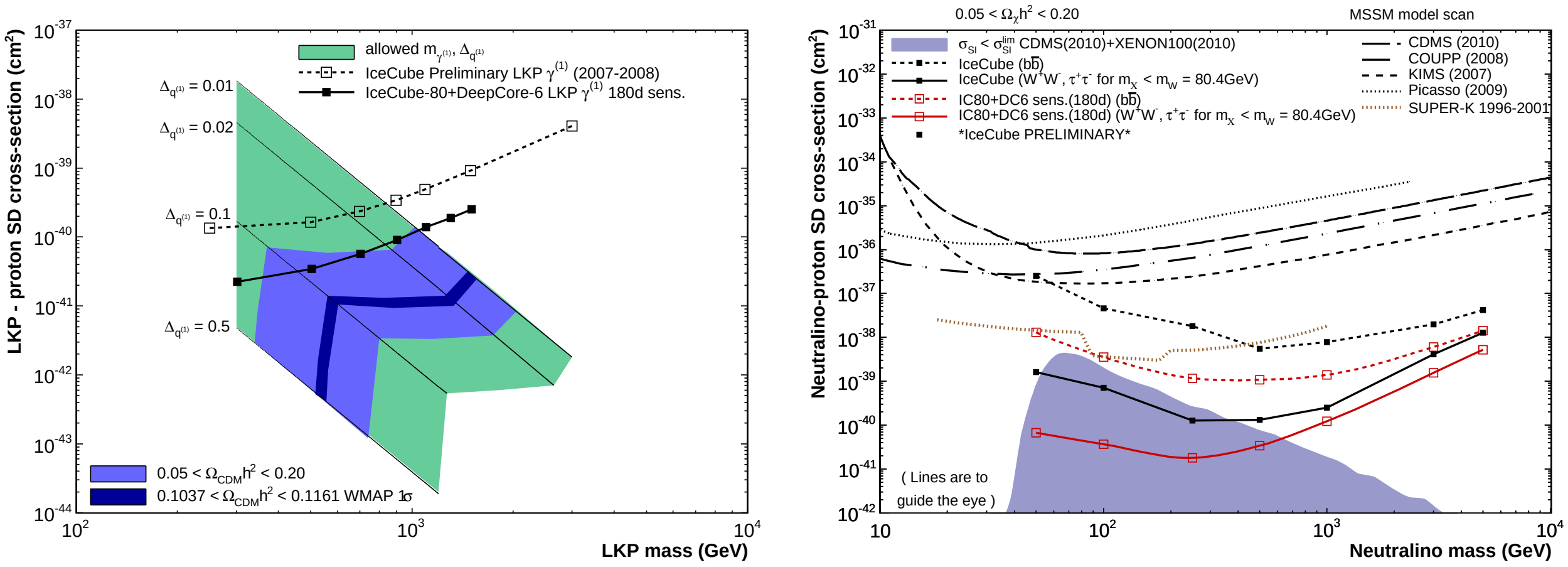

Figure 3: (left plot) Limits on the LKP-proton SD scattering cross-section from IceCube [3] (dashed line) and the final sensitivity for IceCube-86 (solid line) compared to the theoretically allowed region of $m_{\gamma^{(1)}}$ and $\Delta_{q^{(1)}}$. The region below $m_{\gamma^{(1)}} = 300$ GeV is excluded by collider experiments [15]. The upper bound on $m_{\gamma^{(1)}}$ corresponds to the over-closure limit for each individual LKP model [16]. The darker regions (blue in on-line version) indicate the overlap with two different $\Omega_{\rm CDM}h^2$ intervals [17]. (right plot) Sensitivity to $\chi$ on proton spin-dependent cross section at $90\%$ confidence level with respect to $\chi$ mass. IceCube current best limits [3] are shown in solid markers, IceCube-86 sensitivity is shown in non-filled markers and compared to other experiments. Soft WIMP models are indicated by the dashed lines, whereas hard models are shown in solid lines. The indicated model space represents a scan over the allowed MSSM parameter space, accounting for all current experimental constraints.

## 4 Outlook

Construction of the IceCube telescope was completed in December of 2010, creating a sensitive platform for the discovery of Dark Matter candidates in the energy range of 50 GeV to 1 TeV. The co-located DeepCore subarray pushes the energy threshold of IceCube into the 10 GeV regime and allows for the possibility to extend searches to the southern hemisphere. For searches of dark matter annihilation in the center of the Sun, this effectively doubles the experimental livetime while greatly improving the sensitivity for low mass WIMP models where the bulk of the neutrinos and daughter muons are at these very low energies.

We have conducted a detailed simulation-based sensitivity study utilizing the full 86 string detector geometry to better understand the possible physics reach of the detector and have shown that with only 180 days of livetime, we will improve the limits on the spin-dependent WIMP-nucleon scattering cross section by an order of magnitude. In addition, this study shows that the inclusion of the DeepCore detector significantly improves the reach at low WIMP mass, allowing IceCube to probe the full range allowed by current models.

Data-taking with the 79 string configuration of the detector has concluded in May of 2011 and analysis of this dataset has begun. Many improvements in the detector configuration, extraction of useful information from the data, and in analysis techniques for maximizing the WIMP signal lead us to believe that we can substantially improve upon the predictions described in the above sensitivity study. With these improvements in mind, in the coming months we will be able to set the most stringent limits to date on the WIMP-proton spin-dependent scattering cross section, rule out individual models, or potentially detect an excess neutrino flux from the Sun, strongly indicating the capture and annihilation of dark matter in its core.

# Search for Dark Matter in the Milky Way with IceCube

THE ICECUBE COLLABORATION[1]
[1]*See special section in these proceedings*

**Abstract:** We present results of the search for WIMP dark matter accumulated in the galactic halo and the galactic center, using two different configurations of the IceCube neutrino telescope. Limits on the dark matter self-annihilation cross-section at the level of $10^{-22}\,\mathrm{cm}^3\mathrm{s}^{-1}$ – $10^{-23}\,\mathrm{cm}^3\mathrm{s}^{-1}$ are achieved, depending on WIMP mass and the assumed annihilation channel. The status of on-going investigations and future prospects are also discussed.

**Corresponding authors:** Martin Bissok[2] (*martin.bissok@physik.rwth-aachen.de*), David Boersma[2] (*david.boersma@icecube.wisc.edu*), Jan-Patrick Huelss[2], Carsten Rott[3] (*carott@mps.ohio-state.edu*)
[2]*RWTH Aachen, Germany*
[3] *Center for Cosmology and AstroParticle Physics, The Ohio State University, Columbus, OH, USA*

## 1 Introduction

The existence of a cold, non-baryonic dark matter component of the Universe is implied by a variety of astronomical observations. The velocity distribution of galaxy clusters suggests a mass content significantly higher than what should be expected from observation of the luminous mass. On smaller scales, the rotation profiles of galaxies show discrepancies, which hint at the existence of a dark matter halo reaching out far beyond the baryonic disc and bulge. Based on e.g. the anisotropy in the cosmic microwave background observed by WMAP [1], a determination of cosmological parameters is possible, which yields the concordance model of cosmology, with a dark matter content of about 22%. Large scale structure formation and N-body simulation of cosmic evolution require the existence of cold dark matter, which is expected to consist of massive non-relativistic particles. Weakly Interacting Massive Particles (WIMPs) form a class of promising dark matter candidates, that are expected to have masses from a few GeV up to several 100 TeV [2, 3]. Assumed to be relics from the Big Bang they would naturally yield correct cosmic relic abundances. A promising WIMP candidate is the neutralino, the lightest stable particle introduced by the hypothetical minimal super-symmetric extension to the Standard Model.

WIMPs, if assumed to be Majorana particles, can self-annihilate and produce a variety of Standard Model particles in the final state. The dark matter annihilation in the Milky Way would hence yield an excess flux of neutral messenger particles $(\gamma, \nu)$ from the direction of the Galactic Center, where the dark matter density is expected to peak. Further, it would also lead to a large-scale anisotropy over the sky. Such a flux of annihilation products can be detected by earth-based, air-borne or satellite experiments, or the absence of such a flux can be interpreted in terms of upper limits on the velocity-averaged self-annihilation cross-section, $\langle\sigma_A \mathrm{v}\rangle$, of dark matter particles in the considered model. In this paper we present two approaches to the search for dark matter with IceCube; an analysis of a possible large-scale anisotropy and a search for an excess flux from the direction of the Galactic Center.

## 2 Neutrino Flux from Dark Matter Annihilation in the Galactic Halo

Dark matter density profiles based on observations of dark matter dominated galaxies tend to have a rather flat density distribution in the central region, while profiles based on N-body simulations show a steep, often divergent increase of density towards the Galactic Center [4, 5]. An often-used parametrization of (spherically symmetric) dark matter profiles is

$$\rho_{\mathrm{DM}}(r) = \frac{\rho_0}{\left(\frac{r}{r_s}\right)^{\gamma} \cdot \left(1 + \left(\frac{r}{r_s}\right)^{\alpha}\right)^{(\beta-\gamma)/\alpha}} \quad (1)$$

where, $r$ is the distance from the Galactic Center and $r_s$ is the scaling radius. The commonly used NFW-profile is then defined by the parameters: $r_s = 20\,\mathrm{kpc}$, $(\alpha, \beta, \gamma) = (1, 3, 1)$ [6]. $\rho_0$ is chosen so that the locally (at radius of the solar circle $R_{\rm SC}$) assumed dark matter density is $0.3\ {\rm GeV/c^2cm^{-3}}$. Figure 1 illustrates the NFW profile along with two other profiles with different scaling parameters; the flat Kravtsov profile [7] and the cuspy Moore profile [8].

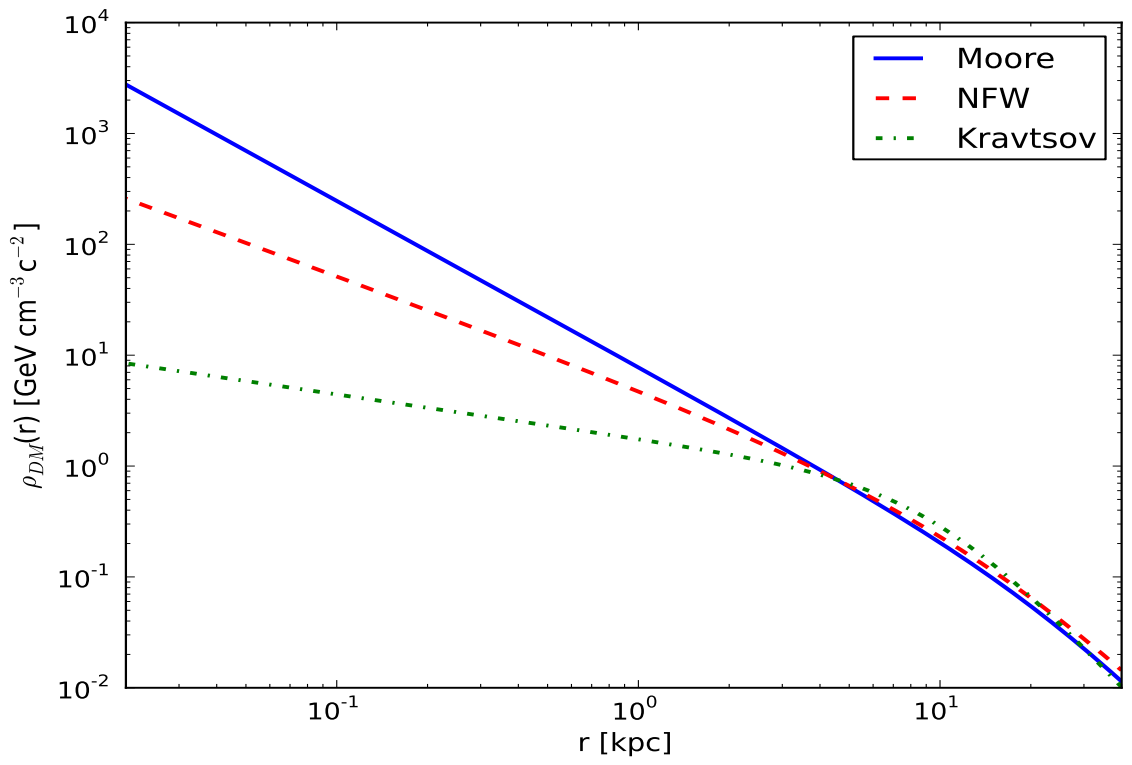

Figure 1: The NFW, Kravtsov, and Moore halo profiles are compared.

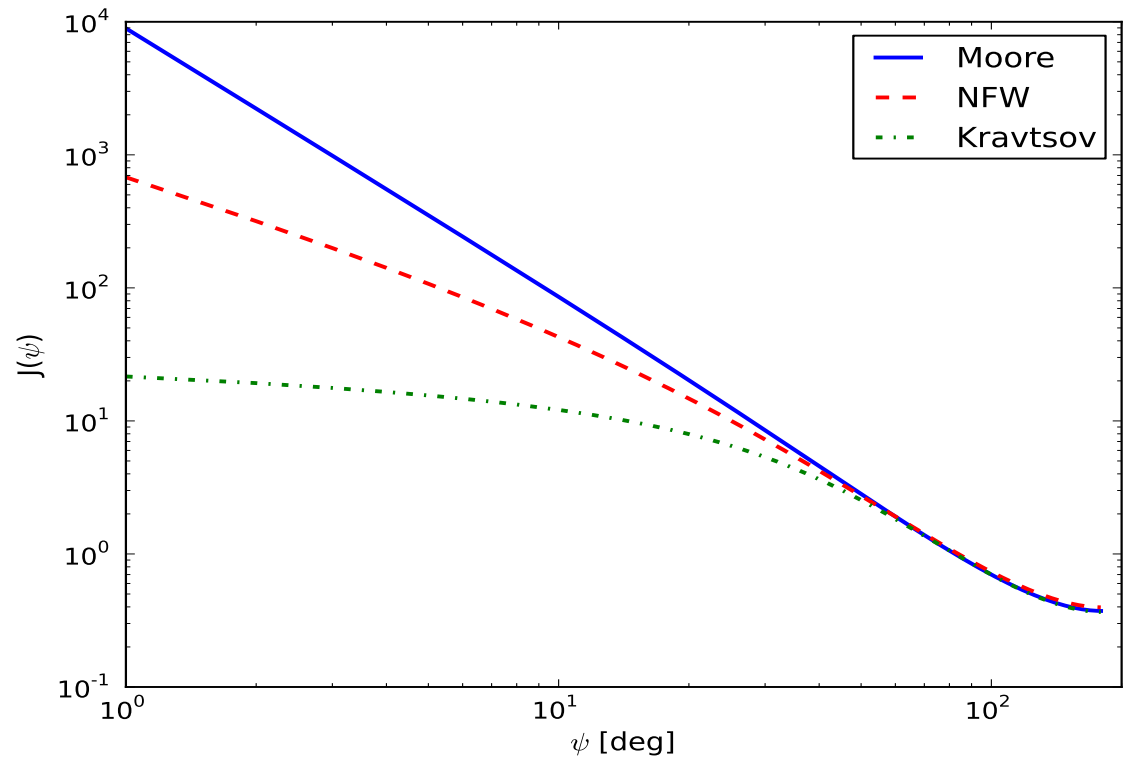

Figure 2: $J(\Psi)$ is shown for the NFW, Moore, and Kravtsov profile.

Following [9], the expected neutrino flux from WIMP annihilation depends on the squared dark matter density integrated along the line of sight

$$J_a(\Psi) = \int_0^{l_{max}} {\rm d}l\ \frac{\rho^2_{\rm DM}(\sqrt{R^2_{\rm SC} - 2lR_{\rm SC}\cos\Psi + l^2})}{R_{\rm SC}\rho^2_{\rm SC}} \quad (2)$$

where $\Psi$ is the opening angle between the line of sight and the Galactic Center and $R_{\rm SC}$ and $\rho^2_{\rm SC}$ are used as scaling parameters, so that $J_a(\Psi)$ is dimensionless. The upper line-of-sight integration limit , $l_{max}$, is chosen to be approximately the Milky Way radius. Figure 2 shows the shape of $J_a(\Psi)$ for the above mentioned profiles. The differential flux can be written as:

$$\frac{d\phi_\nu}{dE} = \frac{\langle\sigma_A v\rangle}{2}\ J_a(\Psi)\ \frac{R_{\rm SC}\rho^2_{\rm SC}}{4\pi m^2_\chi}\ \frac{d\mathcal{N}}{dE}. \quad (3)$$

The factor $1/4\pi$ comes from isotropic emission, $m_\chi$ is the mass of the dark matter particle, and $\frac{d\mathcal{N}}{dE}$ is the energy spectrum of the final state neutrinos. For the presented analyzes, DarkSUSY [10] scans have been performed, for a variety of benchmark WIMP masses and annihilation channels.

## 3 IceCube Neutrino Observatory

IceCube is a km$^3$-size neutrino detector located at the geographic South Pole. Construction has been completed in December 2010. It consists of 86 strings instrumented with 60 Digital Optical Modules (DOMs) each, including a low energy sub-array DeepCore, that has an energy threshold of the order of 10 GeV. The DOMs are located at a depth ranging from 1.45 km to 2.45 km in the Antarctic glacier. The IceCube strings are arranged in a hexagonal symmetry, with an inter-string spacing of 125 m and a DOM distance of 17 m. Eight densely spaced DeepCore strings are deployed between regular IceCube strings and have a DOM spacing of 7 m and an inter-string spacing of 72 m. For the presented analyzes the 22– and 40–string configurations were used, which did not have any DeepCore strings.

## 4 Galactic Halo Analysis with IceCube–22

Between June 2007 and March 2008, the partially-deployed IceCube detector was operated in a configuration with 22 strings and acquired 275 days of live-time, that were searched for neutrino signals from WIMP annihilations in the outer Galactic halo (Northern hemisphere). The sample contained 5114 up-going muon neutrino candidate events, covering $-5°$ and $85°$ in reconstructed declination, with a purity of about 90% [11]. Despite the Galactic Center (RA 17h45m40.04s, Dec -29°00′28.1″) being located on the southern hemisphere and not accessible in this sample, we expect an anisotropy to be present on the northern hemisphere. The anisotropy is caused by the dark matter distribution in the Milky Way halo, which would result in a larger neutrino flux from the region closer to the Galactic Center compared to that further away from it. We have searched for a neutrino anisotropy by comparing fluxes from an on–source region centered around the same RA as the Galactic Center to that from an equally sized off–source shifted by 180° (see figure 3). The analysis was performed in an unbiased fashion, by only counting events in the regions after their size and the analysis procedure were optimized using simulations only.

We observed compatible numbers of events in the on– and off–source region: 1367 and 1389, respectively [12]. As we perform a relative comparison on data the uncertainties on the background estimate can be kept small. A systematic uncertainty of 0.3% due to any pre-existing anisotropy in the data caused by exposure and cosmic-ray anisotropy remains. The signal acceptance uncertainty is approximately 30% and is dominated by the incomplete modeling of ice properties and limitations in the detector simulation. Us-ing equations (2) and (3), a limit on the self-annihilation cross section has been calculated and is shown in figure 4 compared with the limits from the Galactic Center analysis, described in the next section. As the analysis uses the outer halo, the uncertainty on the choice of halo model is small as indicated by the error band on the limits.

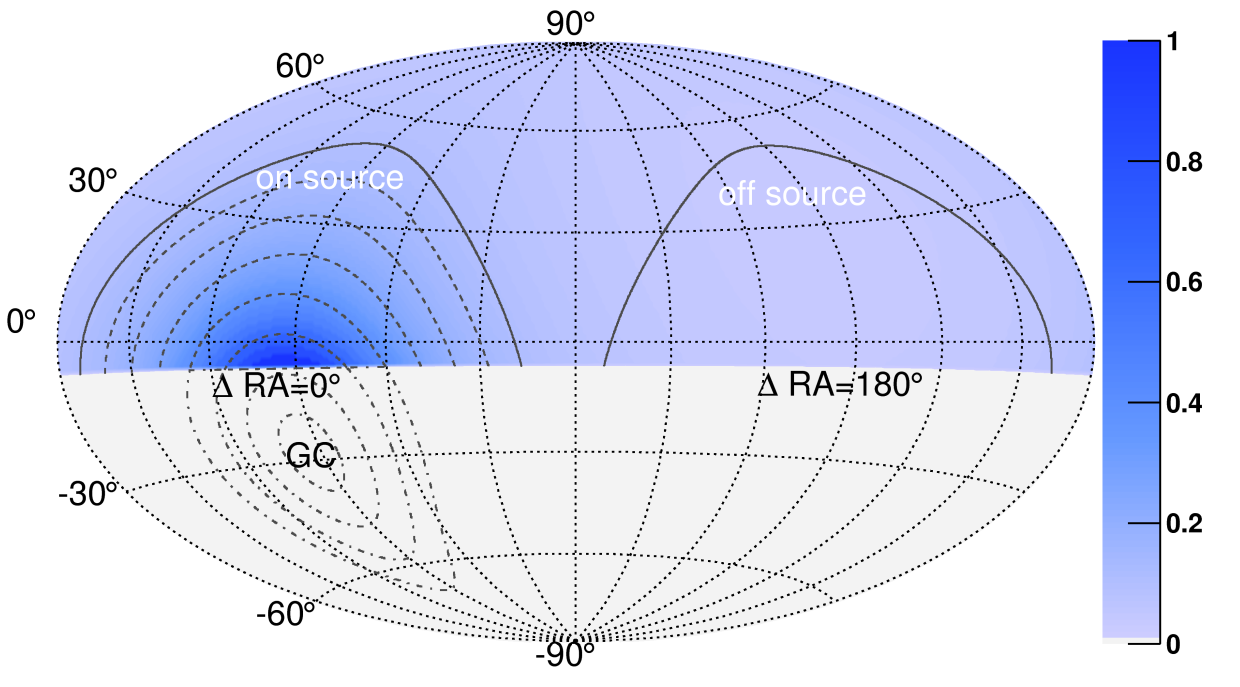

Figure 3: Relative expected neutrino flux in the northern hemisphere from self-annihilation in the Milky Way halo. The on–source region (solid line) is centered around largest the flux expectation at $\Delta\mathrm{RA} = 0$, while the off–source region is shifted by $180°$ in RA.

## 5 Galactic Center Analysis with IceCube–40

The 40–string configuration of IceCube was taking data from April 2008 to May 2009, yielding a total detector livetime of 367 days.

The highest neutrino flux from WIMP annihilation is expected to come from a relatively wide region centered at the direction of the Galactic Center which, at the location of IceCube, is always about 30° above the horizon. Data from this direction is dominated by atmospheric muons, therefore this analysis is based on the identification of events with an interaction vertex inside the detector (atmospheric muons produce incoming tracks) and it relies on the on-source/off-source method; based on Monte Carlo simulations, the width of a declination band (centered at the location of the Galactic Center) is optimized to maximize $\mathrm{signal}/\sqrt{\mathrm{background}}$, assuming the NFW-profile. In this declination band, a window in right ascension is optimized. The optimum window sizes both in right ascension and declination were found to be $\pm 8°$. After correction for uneven exposure, as well as signal quality cuts, the uncertainty on the background prediction is reduced to the $0.1\%$-level. Based on the above mentioned background estimation, the expected number of background events in the signal region was 798819. The number of observed events was 798842. The difference of 23 events is compatible with the null-hypothesis, therefore a $90\%$C.L.-limit on the number of signal events has been calculated (1168), following the Feldman-Cousins approach [13]. Using equations (2) and (3), a limit on the self-annihilation cross-section has been calculated and is shown in figure 4 along with the limits from the previous analysis. Figure 5 shows the obtained limits for the $\tau$ channel, compared to the PAMELA/Fermi regions [14].

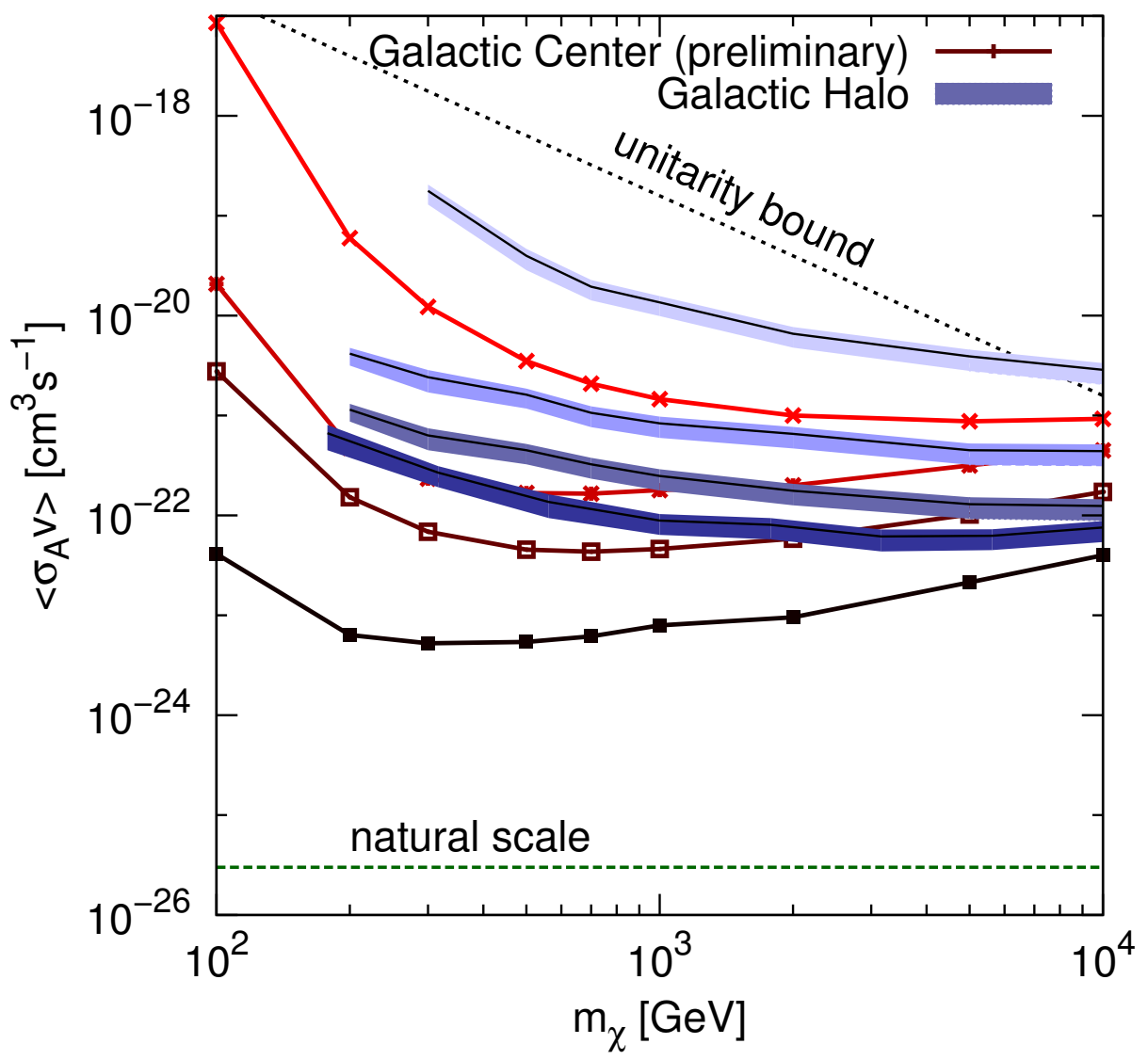

Figure 4: $90\%$ C.L.-limits on the $\langle\sigma_A \mathrm{v}\rangle$ from the IceCube–22 halo analysis (blue-shaded lines) [12], and the limits obtained from the IceCube–40 Galactic Center analysis (simple lines). For both analyses the lines from top to bottom correspond to the $b\bar{b}$, $\mathrm{W^+W^-}$, $\mu^+\mu^-$ and $\nu\bar{\nu}$ annihilation channels. The IceCube–40 limits are preliminary.

The IceCube–40 limits are preliminary, since they do not include signal acceptance systematic uncertainty due to optical ice properties.

## 6 Outlook on the Galactic Center Analysis with IceCube–79

For IceCube–79, a dedicated Galactic Center data filter has been implemented and was taking data from June 2010 to May 2011. The filter consists of two parts. A so-called high energy part accepts all events with a reconstructed arrival direction within an angular window of $\pm 10°$ in declination and $\pm 40°$ in RA with respect to the direction of the Galactic Center and if their brightness exceeds a zenith-dependent threshold. The so-called low-energy part accepts events from a $15°$ wide zenith band around the Galactic Center, but applies a pre-scale factor of 3 on events from the zenith band, which have a distance of more than $20°$ to the Galactic Center in right ascension. Further restrictions for the low energy filter are a top veto defined by the upper 5 DOMs, in which no hits are allowed, and a side veto. The side veto consists of the outer layer of IceCube strings; the earliest pulse is not allowed in this veto region. These filter conditions allow for a preselection of tracks, which appear to start within IceCube. Figure 6 shows a comparison of the effective area at filter level for IceCube–40 and for the low and high-energy filter of IceCube–79. Especially in the low energy region below 100 GeV, more events are accepted. This improvement can be attributed to the Deep-Core array. If these events can be retained throughout the analysis cuts, considerable improvement is to be expected for exclusion limits on the self-annihilation cross-section in the low energy region.

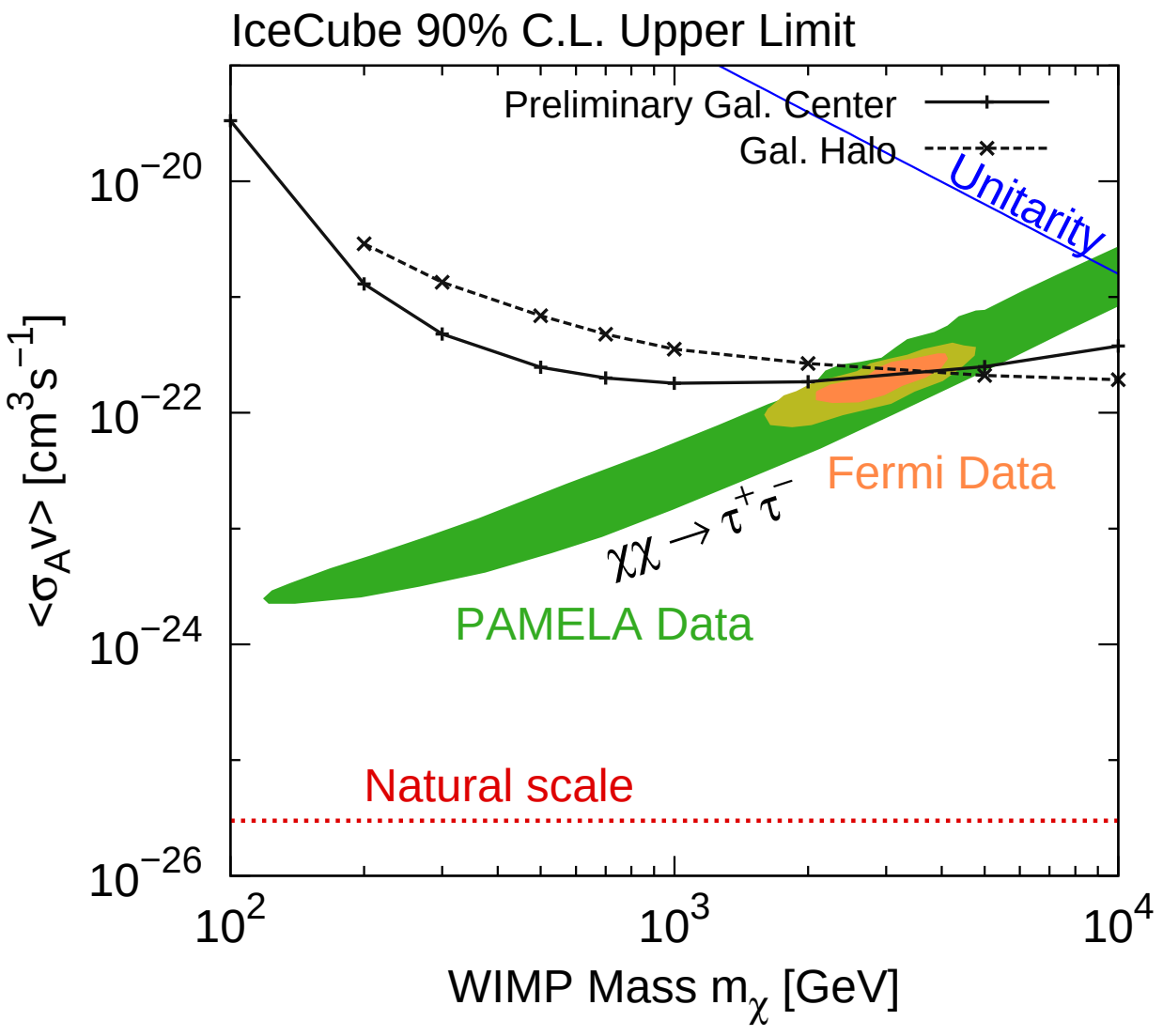

Figure 5: IceCube 90% C.L. upper limits on the $\langle\sigma_A \mathrm{v}\rangle$ from the Galactic halo with the 22–string and the Galactic Center with the 40–string array compared to the preferred regions for PAMELA data, and the region including Fermi data for annihilation to $\tau^+\tau^-$ [14].

## 7 Conclusion

Data collected with the partially instrumented IceCube neutrino detector has been searched for dark matter self-annihilation signals. Two independent analyses, targeting the Galactic halo and Galactic Center, have been performed and resulted in observations consistent with background expectations. Based on these results the dark matter self-annihilation cross section was constrained to $\sim 10^{-22}\mathrm{cm}^3\mathrm{s}^{-1}$ for WIMP masses between 200 GeV and 10 TeV for annihilation into $\tau^+\tau^-$ and $\mu^+\mu^-$. For a neutrino line spectrum $\chi\chi \to \nu\bar{\nu}$, annihilation cross sections larger than $\sim 10^{-23}\mathrm{cm}^3\mathrm{s}^{-1}$ can be excluded, assuming the NFW-profile for the Galactic Center analysis. Limits from the halo analysis are less halo-profile dependent, since the different models show similar behavior for larger distances from the Galactic Center. Despite the small dataset and less than half of the full IceCube detector, the limits already probe a region of interest. A new dedicated filter stream for neutrinos from the Galactic Center has been implemented, that led to an increase in neutrino effective area at filter level of about two orders of magnitude at energies below 100 GeV. With the IceCube detector completed and a dataset available that is already more than three times larger than the ones used for the presented analyzes, we expect to probe dark matter self-annihilation cross sections below $\sim 10^{-24}\mathrm{cm}^3\mathrm{s}^{-1}$. Further, the Galactic halo analysis is currently pursued using the DeepCore detector and the cascade channel ($\nu_e, \nu_\tau$). It utilizes the excellent atmospheric muon veto capabilities with IceCube/DeepCore and lower atmospheric neutrino background in this channel. As the analysis targets a large scale anisotropy, the poor angular resolution of cascade events does not effect this analysis in a strong manner, and will allow for a further improvement in sensitivity.

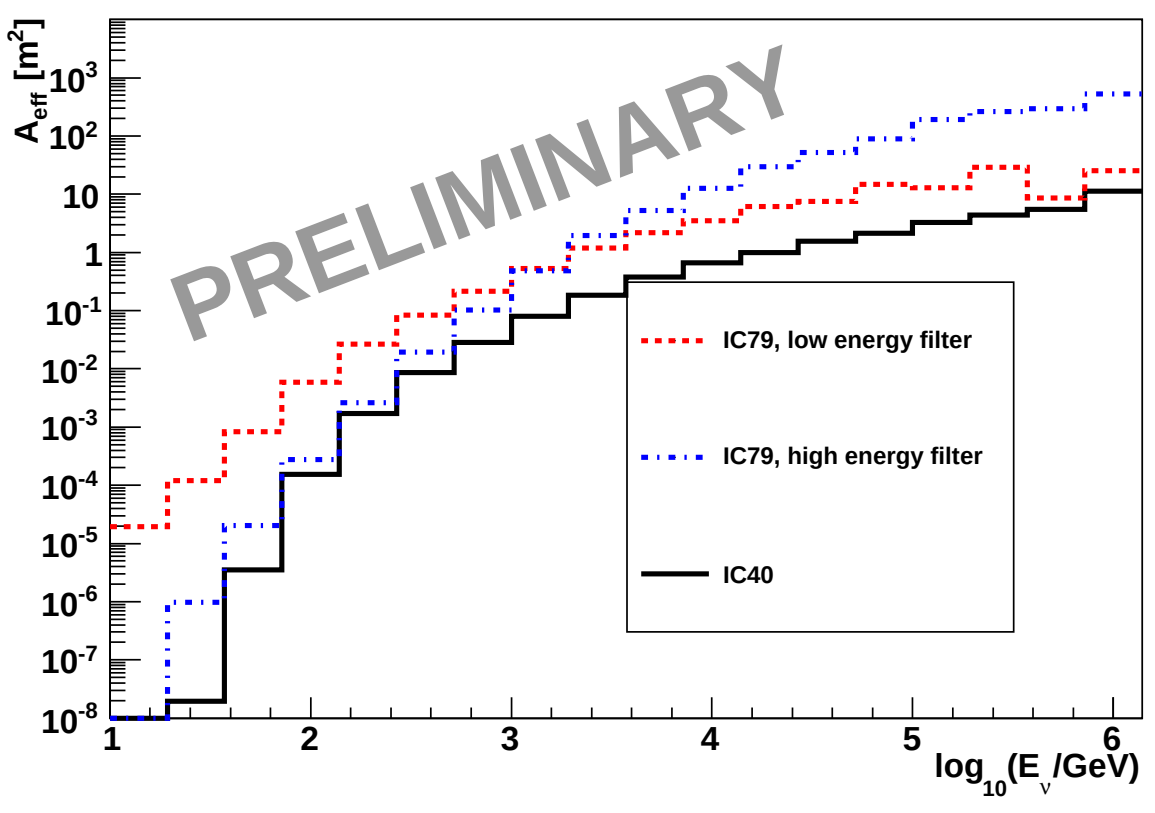

Figure 6: Effective area for IceCube–40, and the two parts of the Galactic Center filter for IceCube–79 at online filter level.

# Search strategies for Dark Matter in nearby Dwarf Spheroidal Galaxies with IceCube

THE ICECUBE COLLABORATION[1]
[1]*See special section in these proceedings*

**Abstract:** Dwarf spheroidal galaxies are believed to contain a large fraction of dark matter due to their high mass-to-light ratio and are therefore promising targets for dark matter searches. They have been investigated by Imaging Air Cherenkov Telescopes and gamma-ray satellites, for which they are excellent targets due to a small field of view and uncomplicated backgrounds. Complementary to such searches, annihilations of WIMPs are also expected to result in neutrino signals that could be detected by the cubic kilometer scale neutrino telescope IceCube. The signal sensitivity can be increased by stacking known dwarf galaxies with the highest flux expectations. We discuss the prospects of the first analysis looking for dark matter in spheroidal galaxies in the northern sky with IceCube, using data taken during 2009 with the 59-string detector configuration.

**Corresponding authors:** Jan Lünemann[2] *(jan.luenemann@uni-mainz.de)*, Carsten Rott[3]*(carott@mps.ohio-state.edu)*
[2]*Johannes Gutenberg University, Mainz, Germany*
[3] *Center for Cosmology and Astro-Particle Physics, The Ohio State University, Columbus, OH, USA*

## 1 Introduction

The observational evidence for the existence of dark matter, from galactic to cosmological scales, is strong. However, its underlying nature remains unknown. A variety of theories provide candidate particles for cold dark matter [1]. Supersymmetry and Extra Dimensions predict new particles with masses around the electro-weak scale, including a stable or long lived weakly interacting massive particle (WIMP) [1, 2]. Such WIMPs form ideal dark matter candidates, whose masses are predicted to be in a range between a few tens of GeV to several TeV.

A variety of standard model messenger particles (including high energy neutrinos) are expected to be produced as a result of the self-annihilation or decay of WIMPs. The neutrinos can be detected by high energy neutrino telescopes, while gamma-rays can be probed with ground based (e.g. H.E.S.S [3]) and space based (e.g. FERMI [4]) telescopes. Electron positron fluxes are sensitive to dark matter annihilations in the vicinity of the Sun ($\leq$ 1 kpc) and have so far resulted in inconclusive results (e.g. PAMELA [5]).

The dark matter self-annihilation cross section can be probed by looking for neutrino signals from the Galactic Center, Galactic halo [6] or dwarf galaxies.

In this paper we discuss the first IceCube search for neutrino signals produced by annihilating dark matter in dwarf galaxies surrounding the Milky Way. These objects are attractive targets for such searches because of their close proximity, relatively compact nature, and their large fraction of dark matter. The search will be used to probe the self-annihilation cross section by constraining the product of cross section and velocity averaged over the dark matter velocity distribution, $\langle \sigma v \rangle$, to probe the lifetime, $\tau$, and to make comparison with recent measurements performed by the FERMI collaboration. Sensitivities will be given for a set of selected benchmark annihilation channels for individual dwarf galaxies.

## 2 Neutrino detection with IceCube

The IceCube Neutrino Observatory [7], located at the geographic South Pole, consists of the IceCube neutrino telescope and the IceTop air shower array. In the ice, a volume of one cubic kilometer of antarctic ice is instrumented with 5160 digital optical modules (DOMs) deployed at depths between 1450 m and 2450 m. The DOMs are distributed over 86 electrical cable bundles that handle power transmission and communication with electronics located on the surface. Each DOM consists of a 25 cm Hamamatsu R7081-02 photomultiplier tube connected to a waveform recording data acquisition circuit.

IceCube is sensitive to all flavors of neutrinos through Cherenkov light emission from secondary particles created when a neutrino interacts in the ice. Muon neutrinos are of particular interest since their extended track-like signature makes them relatively simple to identify and to reconstruct their direction with a few degrees precision at the detection threshold of 50 GeV. This analysis is based on data taken from May 2009 to May 2010 with an intermediate construction stage of the in-ice detector with 3540 DOMs on 59 strings monitoring $\sim 2/3$ of a Gton of antarctic ice. The trigger rate in this configuration is between $\sim 1600$ Hz and $\sim 1900$ Hz and reflects the seasonal modulation of muons produced in cosmic air showers. In future analyses, the denser low energy extension DeepCore, finalized in 2010, will considerably improve the detection of muons with energies below a few hundred GeV. It will also provide sensitivity to celestial objects in the South by employing the veto capacity of the surrounding IceCube strings.

| Source | Right asc. | Declination | Distance (kpc) | Mass ($10^7 M_\odot$) | $J(10^{19} GeV^2/cm^5)$ |
|---|---|---|---|---|---|
| Segue 1 | 10 07 04 | +16 04 55 | 25 | 1.58 | $1.26^{+3.75}_{-0.94}$ |
| Ursa Major II | 08 51 30.0 | +63 07 48 | 32 | 1.09 | $0.58^{+0.91}_{-0.35}$ |
| Willman 1 | 10 49 22.3 | +51 03 04 | 38 | 0.77 | |
| Coma Berenices | 12 26 59 | +23 55 09 | 44 | 0.72 | $0.16^{+0.22}_{-0.08}$ |
| Ursa Minor | 15 09 08.5 | +67 13 21 | 66 | 1.79 | $0.64^{+0.25}_{-0.18}$ |
| Draco | 17 20 12.4 | +57 54 55 | 80 | 1.87 | $1.2^{+0.31}_{-0.25}$ |
| Ursa Major I | 10 34 52.8 | +51 55 12 | 106 | 1.1 | |
| Hercules | 16 31 02 | +12 47 30 | 138 | 0.72 | |
| Canes Venatici II | 12 57 10 | +34 19 15 | 151 | 0.7 | |
| Leo II | 11 13 29.2 | +22 09 17 | 205 | 1.43 | |
| Canes Venatici I | 13 28 03.5 | +33 33 21 | 224 | 1.4 | |
| Leo I | 10 08 27.4 | +12 18 27 | 250 | 1.45 | |
| Leo T | 09 34 53.4 | +17 03 50 | 417 | 1.3 | |

Table 1: List of dwarf spheroidal galaxies ordered by distance from the Earth [14]. J values from [8] and for Segue 1 from [15]. The integral in Eq. 2 is performed over a circle of $0.25°$ radius for Segue 1 and $0.5°$ for the other sources.

The primary background in the search for neutrinos originates from cosmic ray hadronic air showers produced in the Earth's upper atmosphere. The decay of pions and kaons results in a continuous stream of neutrinos and muons. High energy muons are capable of travelling long distances through matter before they eventually decay, resulting in a down-going muon flux at the IceCube detector. We therefore restrict ourselves to dwarf galaxies in the northern hemisphere, using the Earth as a shield.

This paper discusses a sensitivity study of WIMP annihilation in dwarf galaxies. One of the appeals of this analysis is the ability to perform a direct comparison with the search for photons originating in a similar set of dwarf galaxies. We therefore follow the procedures outlined in a recent publication by the FERMI collaboration which set flux limits on several dwarf galaxies [8] and presented a preliminary stacking analysis [9].

## 3 Dwarf galaxies target selection

The number of identified dwarf galaxies has risen in recent years due to a systematic search by the SLOAN Digital Sky Survey (SDSS) [10, 11]. Table 1 lists galaxies in the Northern hemisphere that are accessible to IceCube in the up-going neutrino event sample. The expected neutrino flux from annihilating dark matter is [12]

$$\frac{d\Phi(\Delta\Omega, E)}{dE} = \frac{\langle \sigma v \rangle}{8\pi m_\chi^2} \frac{dN}{dE} J(\Delta\Omega), \qquad (1)$$

where $m_\chi$ denotes the WIMP mass and $\frac{dN}{dE}$ the energy spectrum of the produced neutrinos per annihilation. The "J factor" is the line-of-sight integral of the squared dark matter density:

$$J(\Delta\Omega) = \int_{\Delta\Omega} d\Omega \int_{l.o.s.} \rho_\chi^2(s) ds. \qquad (2)$$

Note that this factor is highly dependent on the assumed dark matter distribution as the annihilation is proportional to the square of the dark matter density $\rho_\chi$. For this paper we assume the commonly used Navarro-Frenk-White (NFW) [13] profile to describe the dark matter density distribution.

## 4 Analysis procedure

In order to cover a large range of possible WIMP signatures, we study WIMP masses in the 0.3–10.0 TeV range and assume a few selected benchmark annihilation channels: $\chi\chi \to \mu^+\mu^-$, $\tau^+\tau^-$, $\mathrm{W^+W^-}$, $\mathrm{b\bar{b}}$, and $\nu\bar{\nu}$. Neutrinos will have undergone extensive mixing through vacuum oscillations over the distances travelled from the dwarf galaxies to the Earth. We determine neutrino flavor oscillations in the long baseline limit. The Monte Carlo results and selection efficiencies are obtained by reweighting the Monte Carlo sample with the expected muon energy distributions for the WIMP masses and annihilation channels studied.

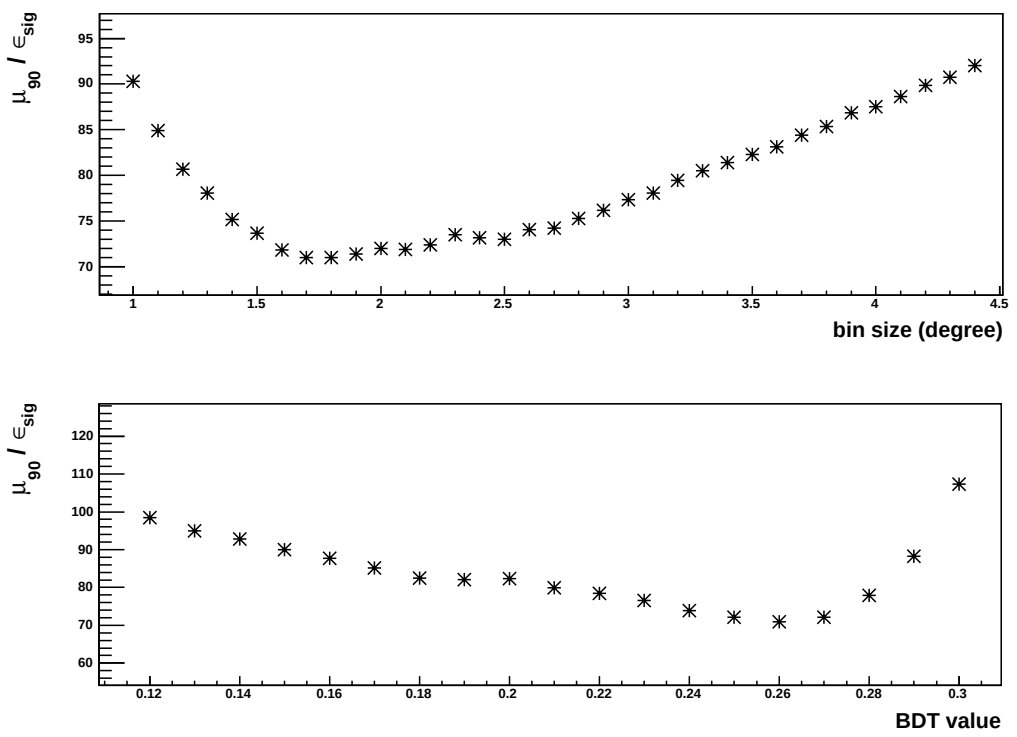

Figure 1: Optimization of selection parameter for Segue 1 and 1 TeV WIMPs annihilating to $\tau^+\tau^-$. Upper plot: Signal mean upper limit ($\mu_{90}$) divided by signal efficiency ($\epsilon_{\text{sig}}$) versus bin size at an optimal cut value of 0.26. Lower plot: Signal mean upper limit divided by signal efficiency versus cut value at an optimal bin size of $1.8°$.

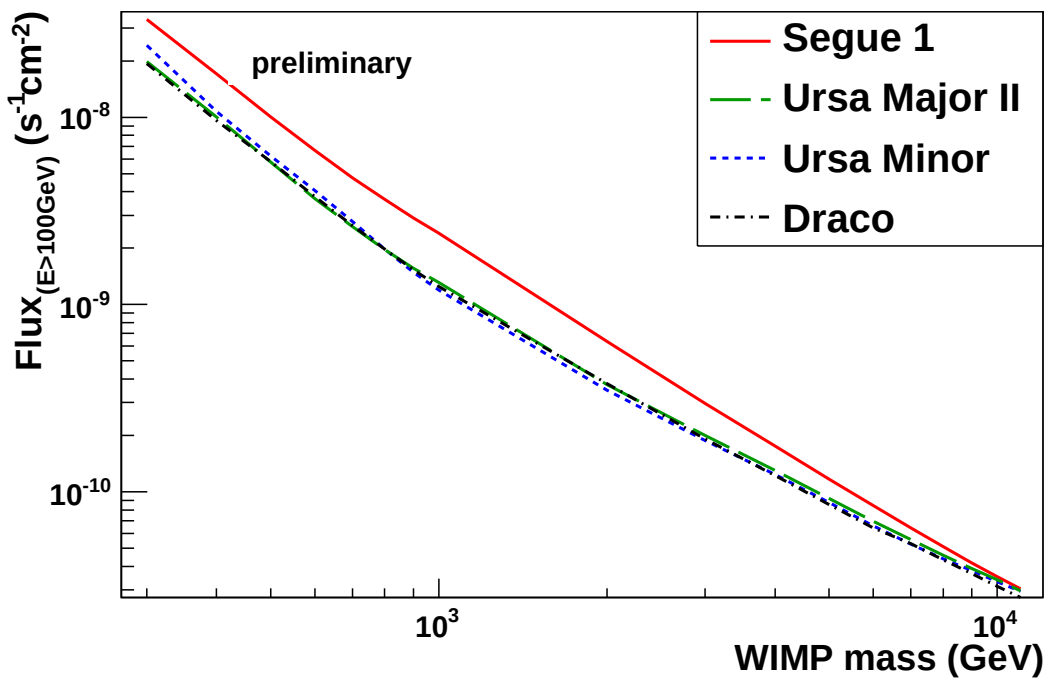

Figure 2: Expected mean upper limit on the neutrino flux versus WIMP mass for different dwarf spheroidals. Values for the $\tau^+\tau^-$ annihilation channel are shown.

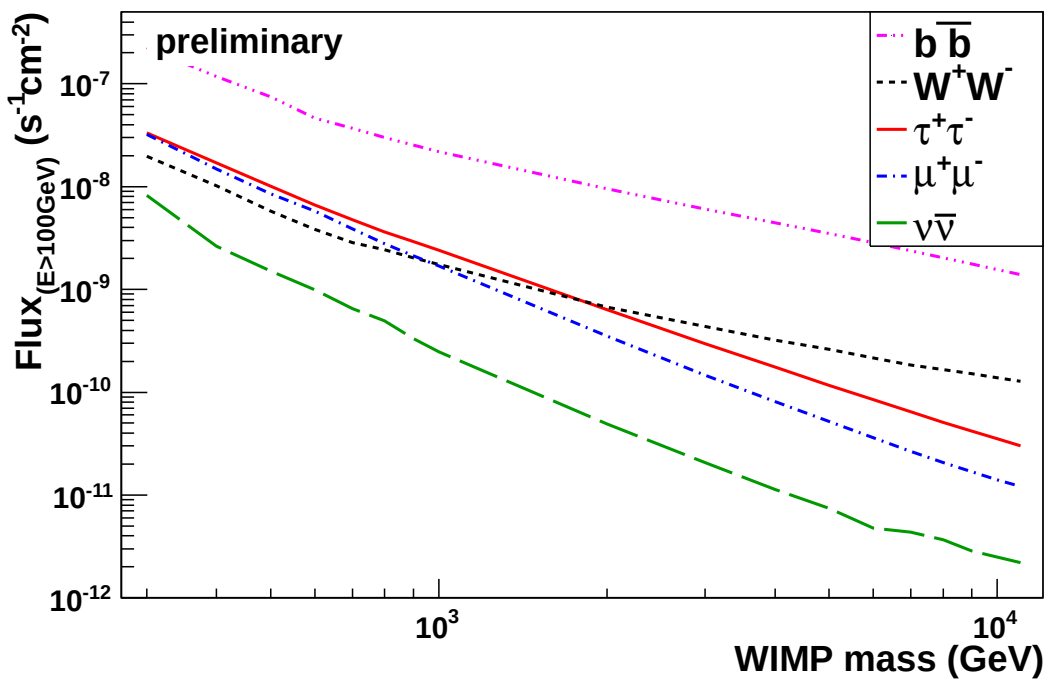

Figure 3: Expected mean upper limit on the neutrino flux versus WIMP mass for different annihilation channels. Values for Segue 1 are shown.

## 5 Event selection and sensitivity determination for individual sources

A pre-selection at the South Pole for up-going reconstructed muon tracks and two further likelihood reconstruction and filtering steps reduces the data rate to $\sim$ 2 Hz. We select events with reconstructed arrival direction 9 degrees below the horizon, as the lowest dwarf galaxy in the North is Leo I at a declination of $12°18'27''$. At this selection cut level the data samples are still dominated by atmospheric muons that are misreconstructed. The sample is then separated into one branch for each WIMP mass and annihilation channel. The event selection is optimized for the expected muon neutrino fluxes obtained with DarkSUSY [16]. Using the Multivariate Analysis package of ROOT (TMVA) several Boosted Decision Trees (BDTs) are trained on simulated signal and background samples, where the signal is weighted to the according energy spectrum for each Tree. The input variables include the likelihood of the track reconstruction, the angular difference between different reconstruction methods, an estimator for the individual angular resolution and the likelihood difference to a reconstruction that is forced to be down-going.

Given the $\sim$ 1 degree angular resolution of IceCube, the dwarf galaxies can be considered point-like. For each WIMP annihilation channel and dwarf galaxy, the event variable cuts and the circular area around the source location are optimized for the most stringent mean upper limit assuming no signal divided by the signal efficiency with respect to the pre-selection. Figure 1 shows the optimization for WIMPs of 1 TeV and assuming self-annihilations into $\tau^+\tau^-$. The background is estimated from a zenith band of $\pm 2.5°$ around the source position. A Feldman-Cousins confidence interval construction [17] is used for the determination of the upper limit. From the corresponding flux a sensitivity for the WIMP annihilation cross section can be derived by using Eq. 1. The average upper limit on the cross section is shown in figure 4.

Figure 2 and 3 show the expected average upper limit on the neutrino flux as function of WIMP mass for different dwarf galaxies and annihilation channels, respectivly. The limits are based on an integrated lifetime of 334.5 days of data-taking with the 59-string detector configuration of IceCube.

## 6 Source stacking

The sensitivity can be improved by analyzing multiple sources simultaneously, a method called source stacking. A prerequisite of this technique is that the flux from all targets differs only in normalization. This is clearly the case for the WIMP annihilation signal of dwarf galaxies. Adding more sources to the analysis lets the signal grow linearly while the background grows proportionally to $\sqrt{N}$. Figure 4 shows how the sensitivity on the annihilation cross

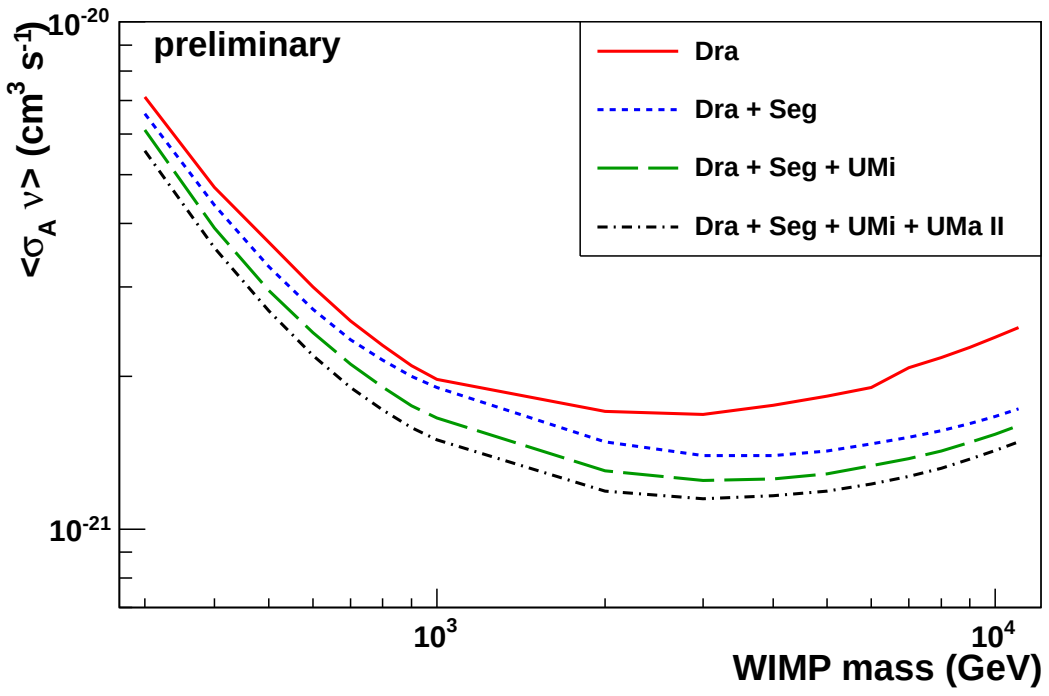

Figure 4: IceCube sensitivities for different number of stacked sources for one year of data in the $\tau^+\tau^-$ channel as function of the WIMP mass.

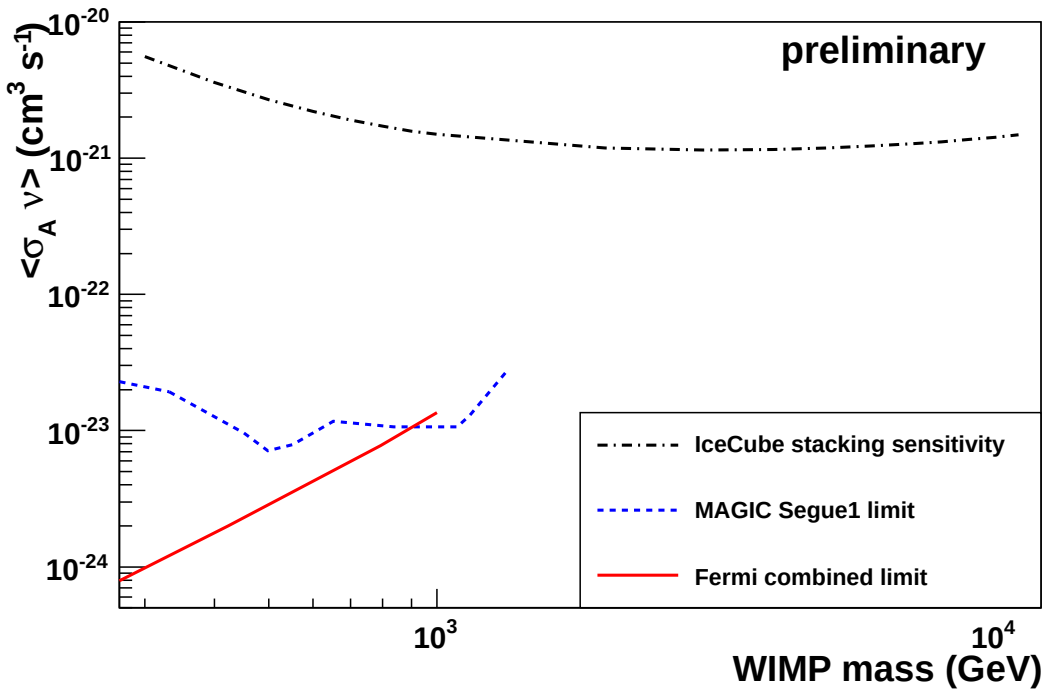

Figure 5: IceCube sensitivity (expected mean upper limit) for one year of data in the $\tau^+\tau^-$ channel as function of the WIMP mass. The sensitivity of the stacking analysis (Draco, Segue, Ursa Minor and Ursa Major II) is compared to results from MAGIC [18] and a preliminary combined limit from Fermi observations of 10 Dwarf Galaxies [9]

section in the $\tau^+\tau^-$ channel improves while increasing the number of stacked sources. In Figure 5 we compare the IceCube sensitivity for four stacked sources assuming dark matter annihilation into $\tau^+\tau^-$, compared to results from MAGIC [18] and a preliminary combined limit from Fermi observations of 10 dwarf galaxies [9].

## 7 Conclusion

We have presented sensitivities for IceCube in the 59-string configuration for the observation of neutrino signals from self-annihilating dark matter in dwarf spheroidal galaxies on the northern hemisphere. Sensitivities for the dark matter self-annihilation cross section are better than $10^{-20}\mathrm{cm}^3\mathrm{s}^{-1}$, for WIMP masses in a range of 300 GeV to several TeV. Searches are complementary to $\gamma$-ray observation in most channels, but have the advantage that IceCube data of these sources is collected continuously and extend to higher WIMP masses. The now operational full IceCube detector will be able to improve the sensitivity to dwarf speroidals further.

# Search Strategies for Relativistic Magnetic Monopoles with the IceCube Neutrino Telescope

THE ICECUBE COLLABORATION[1]
[1]*See special section of these proceedings*

**Abstract:** Several models of unified gauge theory predict massive particles carrying magnetic charge. Galactic or cosmic magnetic fields may accelerate these particles to relativistic velocities. Large scale Cherenkov detectors like IceCube or its predecessor AMANDA are novel tools to search for these particles, since they are predicted to emit several thousand times more light than electrically charged particles. Searches have already been performed with the AMANDA, Baikal and 22-string IceCube neutrino telescopes. We present strategies and methods adopted in the 22 and 40 string detector to separate the expected signal from a background that is several orders of magnitude more abundant.

**Corresponding authors:** Jonas Posselt[2] (*jposselt@uni-wuppertal.de*), Brian Christy[3] (*bchristy@icecube.umd.edu*)
[2]*Dept. of Physics, University of Wuppertal, D-42119 Wuppertal, Germany*
[3]*Dept. of Physics, University of Maryland, College Park, MD 20742, USA*

## 1 Introduction

The existence of magnetic monopoles was first suggested by Pierre Curie in 1894 [1]. However, a firm theoretical grounding was not laid down until 1931 when Paul Dirac [2] demonstrated that magnetic monopoles are consistent with quantum mechanics. Dirac found that magnetic charge is given by $g = Ne/2\alpha \approx 68.5e$, where $\alpha$ is the fine structure constant, and $e$ the fundamental electric charge. In 1974 t'Hooft and Polyakov independently described magnetic monopoles that are regular solutions of the field equations in certain groups of Grand Unified Theories (GUT) and match the charge of the Dirac monopole [3, 4]. Within GUT the masses of magnetic monopoles can be estimated to be $M \propto \Lambda/\alpha$, where $\Lambda$ is the unification energy scale of the theory. This results in a mass range from $10^8$ GeV to $10^{17}$ GeV for various GUT models. Because of these large masses magnetic monopoles are generally assumed to be relics of the early universe where they have been produced via the Kibble mechanism [5].

Analogous to electric charges, which are accelerated along electric field lines, magnetic monopoles are accelerated along magnetic field lines. The kinetic energy gained by a monopole is $E_k \propto gB\xi$, where $B$ is the magnetic field strength, and $\xi$ is the coherence length of the field [6]. During the lifetime of the universe, relic monopoles should have encountered enough accelerators to reach kinetic energies of $\sim 10^{14}$ GeV. Thus monopoles with masses less than $\sim 10^{14}$ GeV should be relativistic. IceCube is able to detect magnetic monopoles travelling through the detector at velocities greater than the Cherenkov threshold ($\beta > 0.76$). The radiation emitted by the monopole is proportional to $(gn)^2$, where $n$ is the index of refraction of the ambient medium [7]. Thus, in ice ($n \approx 1.3$) a monopole will emit $\sim 8000$ times more light than a bare muon of the same velocity.

## 2 Detector

The IceCube detector is a cubic kilometer-scale neutrino telescope. In its now completed state, IceCube consists of 86 strings of 60 Digital Optical Modules (DOMs), each spaced out in a hexagonal pattern and deployed between 1450 to 2450 meters below the Antarctic ice surface. For the data presented, we use the configurations of IceCube as of 2007 and 2008 with 22 and 40 strings respectively. Each DOM is configured to detect and digitize photon signals via a Photomultiplier Tube (PMT) and two waveform digitizers, a fast Analog to Digital Converter (fADC) and an Analog Transient Waveform Digitizer (ATWD) [8]. The ATWD has a sampling rate of $\sim$ 300 Megasamples/second with a total of 128 samples and digitizes the incoming waveform across 3 channels representing different gain values. The fADC runs with a sampling rate of 40 Megasamples/second and can read up to 256 samples. This configuration allows the digitalization of short signals with a high dynamic range as well as long signals, although with a smaller dynamic range.

A DOM is triggered when the PMT signal exceeds a certain threshold. However, the data is only read out in the case of a local coincidence with one neighboring DOM on the same string in a time window of $1\mu$s. This serves to reduce background originating from PMT noise. Since the waveforms produced by a monopole are expected to have a long time scale, the fADC provides greaterdistinction between signal and background. Hence, the search strategies in this study mainly rely on the data provided by the fADC.

## 3 Signal and Background Simulation

The simulation of relativistic magnetic monopoles is done in three stages. Magnetic monopoles are generated uniformly on a disk with their direction perpendicular to the disk. The disk itself is located $\sim$ 1 km from the center of the detector pointing towards it at randomized orientations. The radius of the disk is set to 650 m for the 22 string configuration and 850 m for the 40 string configuration. Datasets were generated for four different speeds, $\beta = 0.995$, $\beta = 0.9$, $\beta = 0.8$, and $\beta = 0.76$, each with isotropic angular distribution at the detector.

Energy loss of the magnetic monopoles as they pass through the ice is modeled using the Bethe-Bloch formula as adapted by Ahlen [9]. The typical energy loss in ice is 6–10 GeV/cm. The light output and propagation is modeled by a version of PHOTONICS [10] specifically generated to work with Cherenkov cone angles associated with the different speeds simulated. The default light amplitude in the simulation is for a muon and this is scaled up using the formula of Tompkins [7].

The principal background to relativistic magnetic monopole searches with IceCube consists of down-going high energy atmospheric muon bundles induced by cosmic rays. Using a 2-component model, which assumes the cosmic ray flux to be composed of only protons and iron, cosmic ray primaries are simulated in the energy range from $10^4$ GeV to $10^{11}$ GeV, where the energy spectra of the components have been fitted to the KASCADE data [11]. The muons are generated with the air-shower simulation package CORSIKA [12] and handed to the detector simulation software.

## 4 Search Strategies

The general procedure used in the two following magnetic monopole searches is that of a blind analysis. Hence the optimization of the data selection is based on simulated data, and only $\sim 10\%$ of the experimental data, referred to as the burn sample, is used for verification purposes. For the monopole signal, a flux of $5 \cdot 10^{-17}$ cm$^{-2}$ s$^{-1}$ str$^{-1}$ is used, roughly representing the lowest limits set by BAIKAL [13] and AMANDA [14].

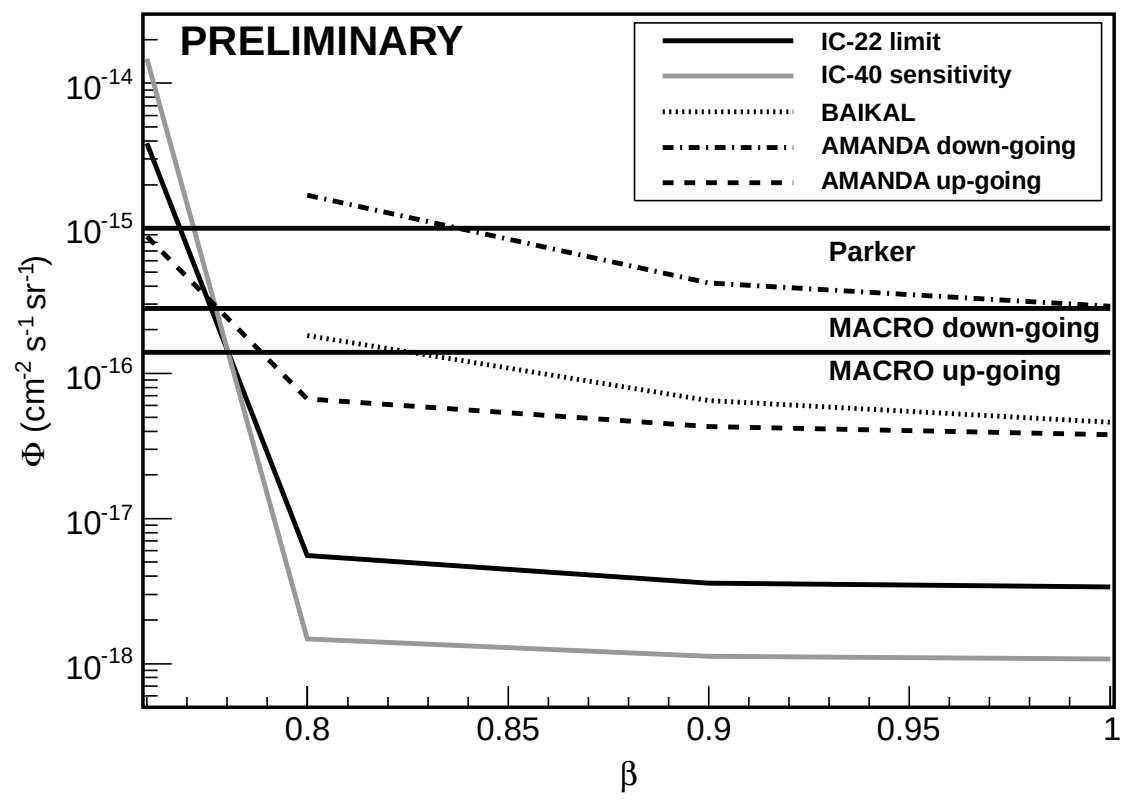

Figure 1: Monopole flux limits set by the 22 string analysis versus beta, shown together with the previous best limits set by AMANDA [14] and BAIKAL [13]. The limits apply to an isotropic flux at the detector. Also shown are the expected sensitivity of the 40 string analysis, limits set by the MACRO [18] experiment and the Parker Bound [19].

### 4.1 IceCube-22

The first search for relativistic magnetic monopoles was performed on the 2007 data run with the detector operating with 22 strings representing a volume of $\sim 0.3$ km$^3$. The selection strategy focused on the brightness of the monopole by first defining a hit with a much higher threshold than the actual DOM, set to be when the fADC waveform reached a saturation level of 1022 ADC counts. This reduced hits from background to be $\sim 10$ m away from the particle track while signal hits were $\sim$ 10–60 m away from the monopole track. Once the sequence of hits was defined, a simple analytic reconstruction was performed based on a plane wave hypothesis and minimizing the variance relative to the actual hit positions, termed ’linefit’. The result of using the higher threshold for a hit leads to an improvement in angular accuracy from 5–10 degrees for all hits down to $\sim 2$ degrees.

The analysis aimed to enhance the sensitivity to slower monopoles by binning the data based on speed reconstruction. Unblinding revealed that the background simulation was not reproducing the tail of the speed distribution where the slower signal was expected to be. This resulted in obvious background events surviving into the final sample. Comparisons based on additional characteristics of waveforms and spatial extent of the hits clearly distinguished the events as background muon bundles. After determining no monopole events were recorded, the following analysis was based only on improved simulation rather than the burn sample or other experimental data. The only changes involved a slight tightening of quality cuts motivated by the new simulation and abandoning any speed binning.

The final cut is in the plane of reconstructed zenith direction and the number of bright hits. For up-going events,

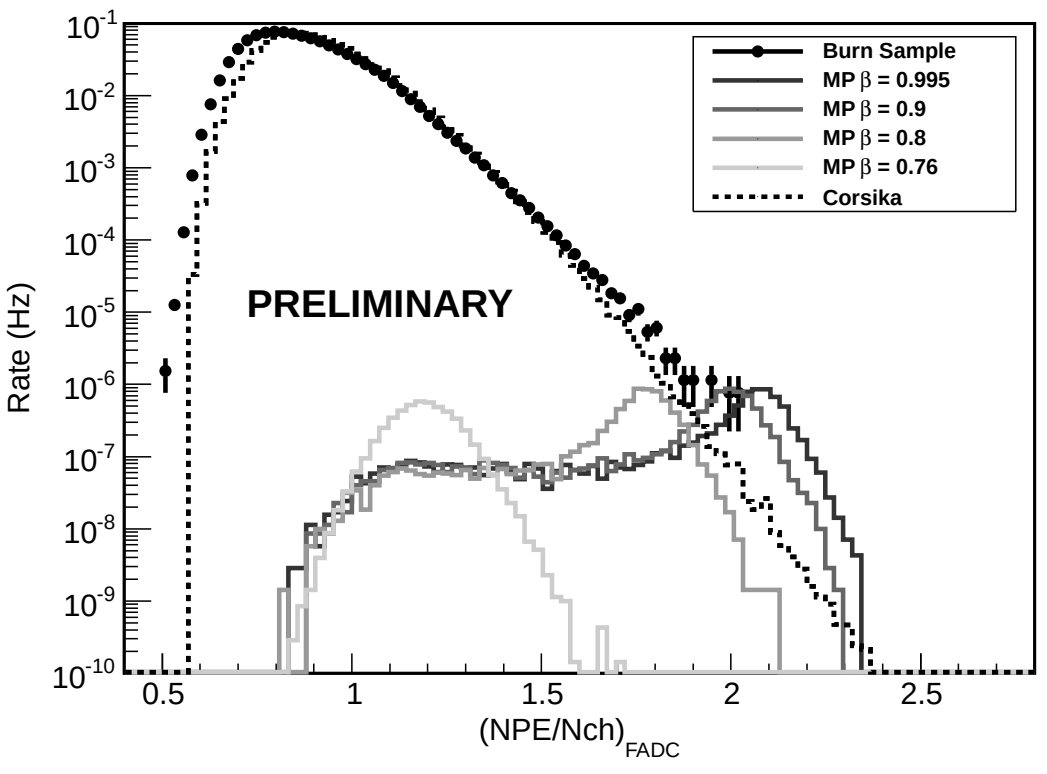

Figure 2: Ratio of the number of hit DOMs and total number of photo-electrons for simulated atmospheric muon background (dashed line), 10% experimental data (black markers) and simulated monopole signal (grey histograms). A precut (Nch $> 50$) has been applied to reduce the discrepancy between experimental data and Monte-Carlo simulation.

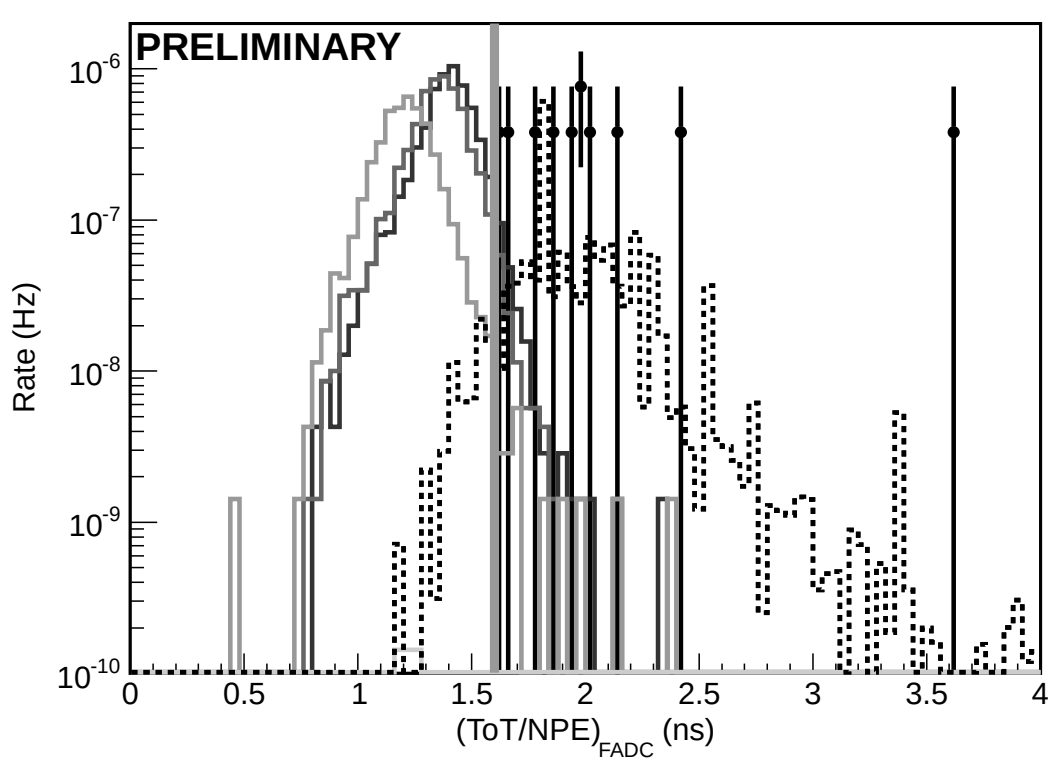

Figure 3: Final cut on time over threshold to NPE ratio. The cut is shown as the thick vertical line at 1.6.

that are largely background free, a simple cut on the number of hits is applied. For down-going directions, where muon bundles produced by high-energy cosmic rays dominate, the cut increases in strength as it approaches the vertical. The cut is optimized on background and signal Monte Carlo by considering the combination that minimizes the model rejection factor [15] for an isotropic and monoenergetic monopole flux.

No data events survived the final cut and upper limits were placed on the flux of magnetic monopoles for speeds of $\beta > 0.8$, which represent a factor of 10 improvement over the AMANDA analysis [16]. Figure 1 shows the limits, as a function of beta, that result for an isotropic flux at the detector. For a complete description of the analysis as well as how these limits transform to an isotropic flux at the Earth's surface, see [17].

## 4.2 IceCube-40

A relativistic magnetic monopole search is also currently underway using the data taken with the 40-string detector. The selection strategy remained focused on the brightness of the monopoles, however, a different observable was chosen. The primary parameter used was the ratio of the total number of photo-electrons (NPE), which is estimated by unfolding the fADC waveforms, and the number of hit DOMs (Nch). While Nch and NPE are in principle integers, their values are large for bright events and therefore the ratio provides a wider range of values than using the number of saturated DOMs. For bright monopoles NPE/Nch is expected to be larger than for atmospheric muon bundles, since the number of DOMs within a radius $r$ around a particle track is proportional to $r^2$, while the number of photons decreases as $\exp(-r/\lambda)$, where $\lambda$ is the effective absorption length. Figure 2 shows the distribution of NPE/Nch after a precut (Nch $> 50$) has been applied to remove events where a particle track outside the instrumented volume triggers the detector. This also significantly improves the agreement of experimental data and Monte-Carlo simulation in the signal region.

The actual cut was made in the plane of reconstructed direction and the inverse of the NPE to Nch ratio, where direction is reconstructed with the linefit method using all hit DOMs. In analogy to the previous search we made a simple cut on the Nch to NPE ratio for up-going events and a cut increasing in strength toward vertical down-going direction. The downside of choosing these cut parameters is that monopoles with $\beta = 0.76$, i.e. close to the Cherenkov threshold, cannot be effectively discriminated from background. The reason for that is the strong decrease of emitted Cherenkov light as the speed approaches the threshold. This narrows the speed interval where the analysis is sensitive, however, only by a small fraction. As an example figure 4 shows the 2-dimensional distribution of the cut parameters for monopoles with $\beta = 0.8$ and simulated background.

The parameter of the final cut is defined as the ratio of the accumulated time the fADC waveforms stay above a certain threshold (ToT) and again the total number of photoelectrons. Figure 3 shows the distribution of the cut parmeter. The cut value is set so that no data events survive the final cut. The estimated sensitivities to magnetic monopoles with speeds $\beta = 0.995$, $\beta = 0.9$ and $\beta = 0.8$ at the detector are given in Table 1 and are also shown in Figure 1. Though no optimization has been applied to the cut values so far, the sensitivity is improved over the analysis using the 22 string detector by a factor of $\sim 4$.

Looking to the future, visual inspection of burn-sample events has revealed two classes of background events which are evading the current cut conditions. The first class is events where a single DOM registers a signal much larger than the remaining DOMs. These events are believed to occur when a particle experiences a stochastic energy loss very close to a DOM. The second and more frequent event class appears to be atmospheric muons with high inclination grazing the corners of the detector resulting in a hit pattern easily leading to misreconstruction. The occurrence of these events may be partially caused by the asymmetry of the 40 string detector. Even though no data events survived the last cut, the final version of the analysis will likely include safety cuts for these event classes.

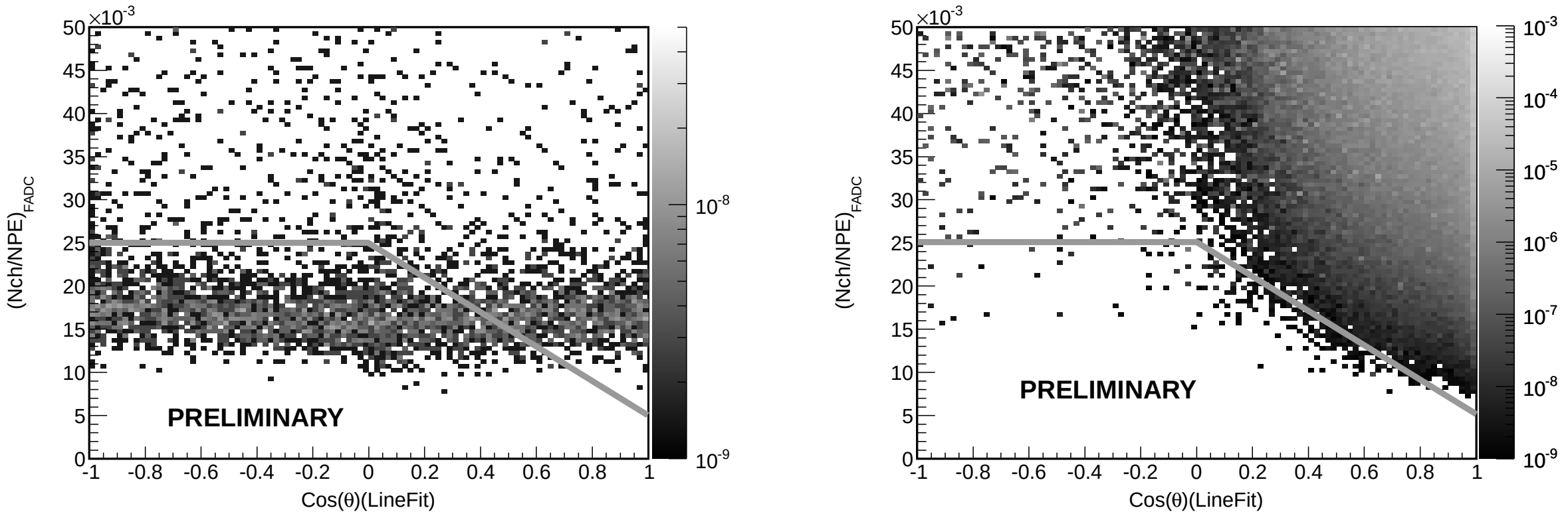

Figure 4: Distribution of Nch/NPE vs Cos($\theta$) for monopoles with $\beta = 0.8$ (left) and for simulated muon bundles (right). A zenith angle of $\theta = 0^\circ$ corresponds to vertically down going. The cut is shown as a grey line, excluding the upper area.

| $\beta$ | $A_{eff}$ km$^2$ | Exp Signal year$^{-1}$ | $\Phi_{90}$ cm$^{-2}$ s$^{-1}$ sr$^{-1}$ |
|---|---|---|---|
| 0.995 | 0.94 | 185.57 | $1.1 \cdot 10^{-18}$ |
| 0.9 | 0.90 | 178.87 | $1.1 \cdot 10^{-18}$ |
| 0.8 | 0.68 | 134.61 | $1.5 \cdot 10^{-18}$ |

Table 1: Effective area, expected signal and sensitivity (90% C.L.) at the detector for a full year of data.

## 5 Conclusion

The analysis performed on the data taken with the 22 string IceCube detector has improved the limits on the flux of magnetic monopoles with speeds $\beta > 0.8$ by an order of magnitude. They are presently the most stringent experimental limits. The ongoing analysis of data taken with the 40 string detector will likely improve these limits by another factor of $\sim 4$. However, these results are preliminary and will be refined.